\documentclass[aps,prl,superscriptaddress,reprint]{revtex4-2}
\usepackage{centernot}
\usepackage{graphicx}
\usepackage{amsmath}
\usepackage{times}
\usepackage{amssymb}
\usepackage{mathrsfs}
\usepackage{chemarr}
\usepackage{xcolor}
\usepackage{url}
\usepackage{version}
\usepackage[normalem]{ulem}

\newcommand{\cL}{\mathcal{L}}

\newcommand{\cO}{\mathcal{O}}

\newcommand{\bu}{\textbf{u}}

\newcommand{\br}{\textbf{r}}

\newcommand{\avg}[1]{\left \langle #1 \right \rangle}
\newcommand{\avgline}[1]{ \langle #1\rangle}

\newcommand{\db}{\partial_{\beta}}

\renewcommand{\rq}{r_{\rm QS}}
\newcommand{\rf}{r_{\rm F}}
\newcommand{\brf}{\bar{r}_{\rm F}}

\newcommand{\rmd}{{\rm d}}

\usepackage[pdftex,colorlinks=true,pdfstartview=FitV,linkcolor= linkcolor,citecolor= linkcolor,urlcolor= linkcolor,hyperindex=true,hyperfigures=false]{hyperref}
\definecolor{linkcolor}{rgb}{0,0,0.6}

\usepackage{lipsum}
\newcommand\bfr{{\bf r}}

\newcommand\bfu{{\bf u}}

\usepackage{tikz}
\usetikzlibrary{calc}
\usetikzlibrary{arrows.meta}
\definecolor{firebrick}{HTML}{B22222}

\newcommand\bfJ{{\bf J}}

\usepackage{hyperref}  
\hypersetup{
    colorlinks=true,
    linkcolor=blue,
    urlcolor=magenta,
    citecolor=red,
    pdftitle={Contact Forces in Motility-Regulated Active Matter},
   	pdfcreationdate={\today},
	pdfmoddate={\today},
    }

\colorlet{shadecolor}{gray!20}

\newcommand{\da}{\partial_{\alpha}}

\newcommand{\dt}{\partial_t}

\newcommand{\dx}{\partial_x}

\renewcommand{\tensor}[1]{\overline{\overline{#1}}}
\renewcommand{\vec}[1]{\bold{#1}}

\newcommand{\vl}{v_{\rm min}}

\newcommand{\rd}{\rho_{\rm d}}

\begin{document}
\let\origaddcontentsline\addcontentsline
\renewcommand{\addcontentsline}[3]{}   

\def\ito/{It\={o}}
\def \pfap/{PFAPs}
\def \qsap/{QSAPs}
\def \assume/{\textbf{ASSUME}}

\title{Contact Forces in Motility-Regulated Active Matter}
\author{{Manh-Quan} Nguyen}
\affiliation{Department of Physics, Massachusetts Institute of Technology, Cambridge, Massachusetts 02139, USA}
\affiliation{Department of Applied Physics, Stanford University, Stanford, California 94305, USA}
\author{Alberto Dinelli}
\affiliation{Department of Biochemistry, University of Geneva, 1211 Geneva, Switzerland}
\affiliation{Universit\'e Paris Cit\'e, Laboratoire Mati\`ere et Syst\`emes Complexes (MSC), UMR 7057 CNRS, F-75205 Paris, France}
\author{Gianmarco Spera}
\affiliation{Rudolf Peierls Centre for Theoretical Physics, University of Oxford, Oxford OX1 3PU, United Kingdom}
\affiliation{Universit\'e Paris Cit\'e, Laboratoire Mati\`ere et Syst\`emes Complexes (MSC), UMR 7057 CNRS, F-75205 Paris, France}
\author{Julien Tailleur}
\email{jgt@mit.edu}
\affiliation{Department of Physics, Massachusetts Institute of Technology, Cambridge, Massachusetts 02139, USA}

\begin{abstract}
Long-range interactions are ubiquitous in nature, where they are mediated by diffusive fields at the cellular scale or by visual cues for groups of animals. Short-range forces, which are paradigmatic in physics, can thus often be neglected when modeling the collective behaviors of biological systems induced by mediated interactions. However, when self-organization leads to the emergence of dense phases, we show that excluded-volume interactions play an important and versatile role. We consider assemblies of active particles that undergo either condensation or phase separation due to motility regulation and show that short-range repulsive forces can induce opposite effects. When motility regulation triggers an absorbing phase transition, such as a chemotactic collapse, repulsive forces oppose the formation of condensates  and stabilize instead the coexistence between finite-density phases. In contrast, when motility regulation induces liquid-gas coexistence, repulsive forces can, counterintuitively, lead to a significant increase in the liquid density. 
\end{abstract}
\maketitle

{Motility regulation refers to a broad set of mechanisms by which active agents adjust their motility, due to interactions or external control. It stems from social forces between pedestrians~\cite{helbing1995social}, aligning torques between birds~\cite{ballerini2008interaction}, directed control~\cite{Fernandez2020,Muinos2021} and communication~\cite{ben2023morphological,zhang2021active} in colloidal and robotic swarms. For microbes, quorum sensing~\cite{miller2001quorum,liu2011sequential,curatolo2020cooperative} and chemotaxis~\cite{budrene1991complex,woodward1995spatio,brenner1998physical,berg2004coli,zhao2023chemotactic,zhao2026dynamic} are established motility-regulation pathways that drive ecosystem self-organization. Interestingly, they are also relevant for colloids where they emerge spontaneously~\cite{Geyer2019,zhang2021active,lefranc2025quorum} or through engineering~\cite{bauerle2018self,lavergne2019group}.}

The simplest form of {collective} organization induced by motility regulation is arguably {scalar} phase separation, {which is characterized by a single hydrodynamic mode, the density field. Phase separation emerges in biological and synthetic matter from a wide range of motility-regulation mechanisms, such as chemotaxis~\cite{brenner1998physical,o2020lamellar,zhao2023chemotactic}, 
quorum-sensing~\cite{tailleur2008statistical,liu2011sequential,bauerle2018self,curatolo2020cooperative,lefranc2025quorum}, 
or diffusiophoretic forces~\cite{palacci2013living,theurkauff2012dynamic,soto2014run,pohl2014dynamic,saha2014clusters}. In these cases, motility regulation is mediated, e.g.~by diffusive signaling molecules. It is thus already relevant in dilute systems and contact forces
are typically neglected in the corresponding theoretical
models~\cite{tailleur2008statistical,saha2014clusters,cates2015motility,Solon_2018_Rmap,gnan2022critical,ridgway2023motility,dinelli2023non,duan2023dynamical}. 

However, this approximation  becomes questionable once dense phases arise from phase separation. 
This is particularly concerning when motility regulation
triggers real-space
condensation~\cite{brenner1998physical,evans2005nonequilibrium,tailleur2008statistical,saha2014clusters,golestanian2019bose,o2020lamellar,mahault2020bose}, leading all particles to aggregate into dense condensates in which contact
forces cannot be ignored. 
Furthermore, it has been
suggested numerically that the interplay between motility regulation and pairwise
forces is a robust self-organization
pathway~\cite{peruani2008dynamics,soto2014self,paoluzzi2018fractal,abaurrea2018collective,paoluzzi2020information},
 likely to play an important role in dense living systems
like bacterial
colonies~\cite{saintillan2008instabilities,baskaran2009statistical,marchetti2013hydrodynamics}
and epithelia~\cite{tlili2018collective,alert2020physical}. Indeed, contact interactions can themselves trigger self organization in biological~\cite{mayor2010keeping,liu2019self} and synthetic~\cite{buttinoni2013dynamical,Geyer2019,lefranc2025quorum} active matter.
However, we still lack a theoretical framework to predict the collective behaviors that emerge from the joint action of motility regulation and pairwise forces.}

\begin{figure}
    \centering
    \includegraphics{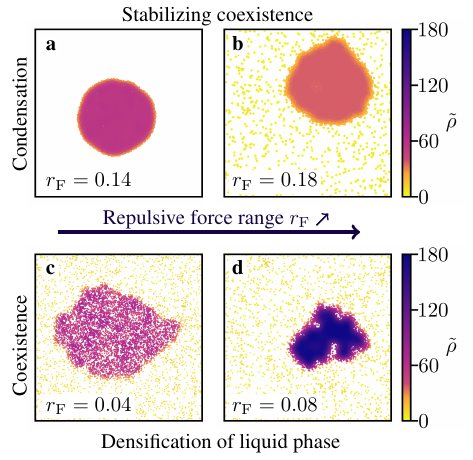}
    \caption{Impact of repulsive forces of range $\rf$ on the self-organization induced by quorum-sensing interactions. The latter are modeled using a decreasing density-dependent self-propulsion speed, $v(\tilde \rho)$, where the local density $\tilde \rho(\br)$ is averaged over a radius $\rq=1$.
    {\bf (a-b)} When $v(\rho\to\infty)= \vl=0$, quorum sensing induces a condensation transition into fully arrested droplets (a). Repulsive forces then oppose the arrest and stabilize a liquid-gas coexistence (b). {\bf (c-d)} When $\vl$ is small but finite, quorum sensing induces a liquid-gas coexistence (c). Repulsive forces are then shown to induce a fourfold densification of the liquid phase (d). All panels correspond to simulations of Eq.~\eqref{eq:dynamics}.
    }
    \label{fig:Fig1}
\end{figure}

In this {Letter}, we make progress on this question by considering self-propelled particles that interact via pairwise forces and undergo a phase separation induced by motility regulation.
{For the sake of concreteness, we focus on the case where the particle speed is regulated by quorum sensing,} which has attracted a lot of attention recently~\cite{tailleur2008statistical,Solon_2018_Rmap,gnan2022critical,ridgway2023motility,dinelli2023non,duan2023dynamical}. 
Our central result is that repulsive forces play a versatile role in phase-separating motility-regulated systems, as illustrated in Fig.~\ref{fig:Fig1}.
First, when motility regulation induces a condensation transition leading all particles to be absorbed in nonmotile droplets, repulsive forces oppose condensation and instead stabilize a \textit{bona fide} liquid-gas coexistence. 
In contrast, if motility regulation leads to liquid-gas coexistence, repulsive forces can induce a severe densification of the liquid phase. 
Below, we detail these numerical results and develop a theoretical framework for predicting the phase equilibria of self-propelled particles {interacting via quorum sensing in the presence of pairwise forces. While part of our theory is specific to this system, our results are not: In the End Matter, we show that chemotactic interactions lead to the exact same phenomenology as in Fig.~\ref{fig:Fig1}.} We believe that our results should have implications both for biological systems, where mediated interactions are ubiquitous, and for the design and control of synthetic active materials.

\textit{Model.} We consider $N$ active Brownian particles in a $2d$ box of size $L\times L$ with periodic boundary conditions. Their 
 positions $\bfr_i$ and orientations $\bfu_i = (\cos \theta_i, \sin \theta_i)$ evolve as
\begin{equation}\label{eq:dynamics}
\dot{\vec{r}}_i = v[\tilde\rho(\bfr_i)] \vec{u}_i - \sum_j \vec{\nabla} V(\vec{r}_{i}-\vec r_{j})\;, \quad
\dot{\theta}_i = \sqrt{2D_r} \eta_i\;.
\end{equation}
Here, $D_r$ is the rotational diffusivity, $\eta_i$ is a centered Gaussian white noise with unit variance, and the time unit is such that the particle mobility equals $1$.  {All functions, parameters, and simulation details are given in the End Matter.}

{Motility regulation is implemented through the quorum-sensing (QS) interaction $v(\tilde \rho)$}, where $v$ is a sigmoidal function decreasing from  $v_{\rm max}$ to $v_{ \rm min}$ at a typical threshold density $\rho_{\rm t}$, which models the response of quorum-sensing circuits~\cite{liu2011sequential}.
The local density is sampled as $\tilde\rho(\bfr)=\sum_i K(\bfr-\bfr_i)$, with $K$ a normalized kernel of range $\rq$. {Soft pairwise forces (PFs) are accounted for via a  repulsive potential $V(r)$ of range $\rf$. 
Note that, in the absence of pairwise forces, the slowing down at high density induced by quorum sensing suffices to trigger a motility-induced phase separation (MIPS), leading to the coexistence between a dilute active gas and a dense, slower liquid phase~\cite{tailleur2008statistical,Solon_2015_ABP_RTP}. We refer to this phase transition as QS-MIPS.
Conversely, in the absence of quorum sensing, the slowing down induced by repulsive pairwise forces is also able to trigger MIPS at large enough speed and packing fraction~\cite{Fily_2012_origin,redner2013,stenhammar2014phase,wysocki2014cooperative,cates2015motility,caporusso2020motility,Speck_2021,Omar_2023}, which we refer to as PF-MIPS. 

For a system at density $\rho_0=N/L^2$, several lengths are at play: the interaction ranges $\rq$ and $\rf$, the persistence length $\ell_p =v/D_r$, the  interparticle separation $\rho_0^{-1/2}$, and the quorum-sensing threshold $\rho_{\rm t}^{-1/2}$. Their roles in the transitions shown in Fig.~\ref{fig:Fig1} will be discussed below. Already, we take $\rf\leq r_{\rm QS}$ to model long-range motility regulation by quorum sensing and short-range repulsion, and we set the unit of length such that $\rq=1$. 
 }

\bigskip

\begin{figure}[t!]
  \includegraphics[width=1.0\columnwidth,  trim = {0 0 7pt 0}, clip]{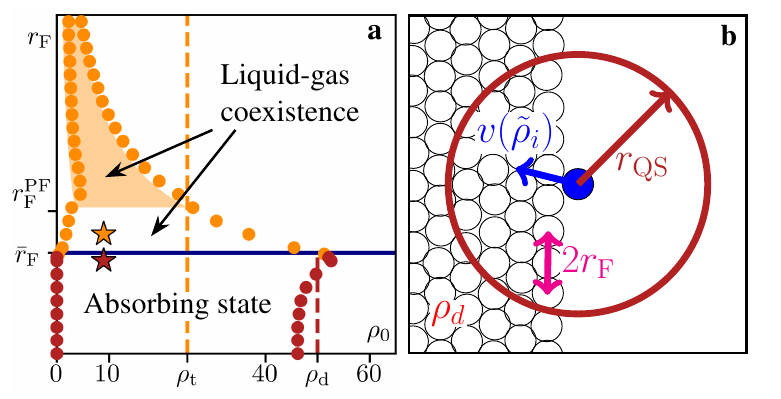}
    \caption{({\bf a}) Phase diagram in the $(\rho_0, \rf)$ plane when quorum sensing induces an absorbing phase transition into an arrested state due to $v(\rho>\rho_{\rm t})=0$.
    Red and orange stars correspond to simulations shown in Fig.~\ref{fig:Fig1}a-b, respectively. For $\rf<\brf \approx 0.15$, the system always reaches the absorbing state. At large densities, $\rho>\rho_{\rm t}$, homogeneous configurations are trivially arrested. At lower densities, the system undergoes a condensation transition, leading to the coexistence between an empty gas and arrested droplets, whose densities are marked with red disks. The solid blue line represents the prediction~\eqref{eq:brf} for $\brf$. The dashed red line is the theoretical prediction for the droplet density $\rd = 2\rho_{\rm t}$. For $\rf>\brf$, we observe the coexistence between liquid and gas phases, whose densities are marked with orange disks. The shaded region corresponds to the MIPS binodal predicted in the absence of quorum-sensing interactions~\cite{Solon_2018_Rmap,Omar_2023,supp}, valid down to $\rf^{\rm PF} \approx 0.24$ where the predicted liquid binodal equals $\rho_{\rm t}$.  
    ({\bf b}) Sketch of the absorption of an active particle at the interface of a static cluster.}
    \label{fig:Fig2}
\end{figure}

\textit{Repulsive forces oppose condensation and arrest.}
When quorum sensing  leads to a vanishing motility at high density [$v=0$ for $\rho\geq \rho_{\rm t}$], the system undergoes an absorbing transition into an arrested state in the absence of repulsive forces ($\rf=0$). Starting from a homogeneous system, all particles condense into arrested droplets (See  Figs.~\ref{fig:Fig1}a-b and SM Movies~1-2~\cite{supp}). As  $\rf$ increases, the droplets first densify until a critical value $\brf$ beyond which the system transitions out of the absorbing phase: the arrested state is then replaced by an ergodic phase illustrated in Fig.~\ref{fig:Fig1}b, where a dilute gas coexists with a dense liquid. Let us now show how to account for the corresponding phase diagram, reported in Fig.~\ref{fig:Fig2}a. 

We first consider the absorbed state, whose existence can be understood qualitatively at {the} mean-field level at $\rf =0$. In $d$ dimensions, standard methods~\cite{tailleur2008statistical,Cates_2013_ABP_RTP} allow approximating the large-scale dynamics of active particles as $\dot \rho=-\nabla \cdot \mathbf{J}^{\rm QS}$, where the particle current reads $\mathbf{J}^{\rm QS} = -D \rho \nabla \mu^{\rm QS}$ with $D=v^2/[d(d-1) D_r]$. 
{Here, $\mu^{\rm QS} = \log (\rho v)$ plays a role akin to that of a chemical potential in equilibrium model-B dynamics~\cite{hohenberg1977theory}. The nucleation of an arrested droplet at density $\rd>\rho_{\rm t}$ leads to $v(\rd) = 0$, which sends the droplet effective chemical potential to $\mu^{\rm QS}=-\infty$. 
This
results in a particle current $\mathbf{J}^{\rm QS}$ that always points toward the droplet, leading to the progressive absorption of the entire system. 
Note that this coarse-grained picture applies beyond the system at hand, for instance to account for the absorbing arrested phase observed in Quincke rods~\cite{lefranc2025quorum}.} 

{However, this mean-field approach does not allow for the prediction of} the droplet density $\rd$, nor its fate as $\rf>0$, and we thus complement the above reasoning with a kinetic argument.
For static clusters to grow, incoming particles need to detect a local density $\tilde \rho(\bfr_i) > \rho_{\rm t}$ at the cluster boundary. 
There, as sketched in Fig.~\ref{fig:Fig2}b, the space is divided between the droplet at density $\rd$ and a dilute gas with $\rho \approx 0$. 
Arrest thus occurs when $\tilde\rho(\bfr_i) \approx \rd/2=\rho_{\rm t}$, leading to $\rd \approx 2 \rho_{\rm t}$. 
{As shown in Fig.~\ref{fig:Fig2}a, all our simulations  up to the transition value $\rf=\brf$ fall within $10\%$ of the prediction $ \rd=2\rho_{\rm t}$. In~\cite{supp}, we show how the coarsening dynamics and repulsive interactions lead to corrections to this prediction that explain the small variations of $\rd$ with $\rf$.}

As $\rf$ increases beyond the critical value $\brf$, the system enters an ergodic
phase. The absorbing droplets are then suddenly replaced by the stable coexistence between liquid and gas phases, as illustrated in Fig.~\ref{fig:Fig1}b. To understand this transition, we note that each particle occupies an effective area $\pi \rf^2/4$. The maximal density allowed for spheres in $2d$ is {$\rho^{\rm max} = 4\phi^{\rm hex} / (\pi\rf^2)$, where $\phi^{\rm hex} = \pi/(2\sqrt{3})$} is the  packing fraction of a hexagonal lattice of disks. Our kinetic argument thus predicts that the absorbing phase ceases to exist when $\rho^{\rm max}<\rd=2\rho_{\rm t}$, leading to the prediction 
\begin{equation}\label{eq:brf}
\brf = (\sqrt{3}\rho_{\rm t})^{-1/2}\;.
\end{equation}
As shown in Fig.~\ref{fig:Fig2}a, this prediction successfully captures the transition observed in our simulations{, revealing that the ratio between the two length scales $\rf$ and ${\rho_{\rm t}}^{-1/2}$ controls the transition into the absorbing phase.}

As $\rf$ increases beyond $\brf$, the role of {quorum sensing becomes less and less relevant until it plays no role for $\rf>\rq$}. 
In the absence of quorum sensing, MIPS is solely controlled by the pressure resulting from interparticle forces and self-propulsion~\cite{Solon_2015_pressure_ABP}. The coexisting densities are thus determined by the packing fraction $\phi \sim (\rf \sqrt{\rho_0})^2$ and the persistence length scaled by the particle size $\sim \ell_p / \rf$~\cite{Solon_2018_Rmap,Speck_2021,Omar_2023}. We denote by $\rho_{\ell}^{\rm PF}$ and $\rho_{g}^{\rm PF}$ the corresponding liquid and gas binodals. 
We predict that neglecting quorum sensing is justified as long as {the liquid density lies below the threshold for motility regulation, i.e.}~$\rho_{\ell}^{\rm PF} \leq \rho_{\rm t}$, {so that} quorum sensing barely alters $v(\rho) \simeq v_{\rm max}$. This condition corresponds to $\rf= \rf^{\rm PF}\simeq 0.24 \,\rq$, which agrees well with the results shown in Fig.~\ref{fig:Fig2}a. {We show in~\cite{supp} that our predictions for the phase diagram are robust to the precise values of $r_{\rm QS}$ and $\rho_{\rm t}$.}

\bigskip

\begin{figure}
    \centering
    {\includegraphics[width=\columnwidth, trim = {7pt 0 0 0}, clip]{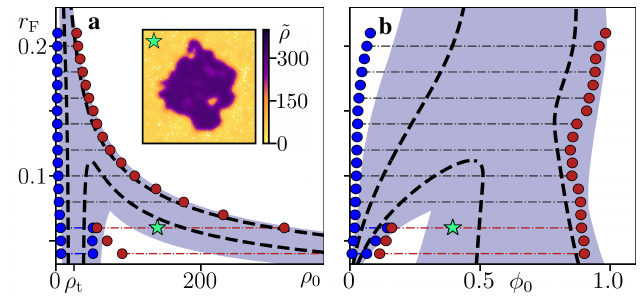}}
    
    \caption{Phase diagram in the  $(\rho_0,\rf)$ plane {(a)} and in the $(\phi_0,\rf)$ plane {(b)} when $v(\rho\to\infty)> 0$. Disks connected by dash-dotted lines correspond to the measured coexisting binodals. For $\rf \leq 0.06$, QS-MIPS and PF-MIPS binodals are depicted as blue and red disks, respectively. Dashed black lines indicate the spinodals predicted by Eq.~\eqref{eq:linear_stability}. The shaded region indicates the binodals predicted from the local theory~\eqref{eq:local_theory}. As expected, predicted and measured binodals agree qualitatively (without free parameters) but not quantitatively, due to the mean-field approximation and the neglected gradient terms. (See~\cite{supp} for a refined theory.) The inset of panel (a) illustrates the PF-MIPS at $\rho_0=140, \rf=0.06$ (green star).
    }
    \label{fig:phase_diag_densification}
\end{figure}

\textit{Pairwise forces densify the liquid phase.}
We now consider quorum-sensing  interactions such that the self-propulsion speed remains finite at large densities: $v_{\rm min} {=  v(\rho \to \infty)} > 0$. In the absence of steric repulsion ($\rf=0$), {quorum sensing} induces MIPS between a slow liquid and a dilute active gas~\cite{tailleur2008statistical,cates2015motility}.  As {the repulsion range} $\rf$ increases, pairwise forces come into play and  lead to a counterintuitive densification of the active phase at {$\rf \simeq 0.06$} (Figs.~\ref{fig:Fig1}c-d and SM Movies~3-4~\cite{supp}). {As we now show, this results from the interplay between the two competing mechanisms that can induce MIPS. First, the quorum-sensing-driven instability, QS-MIPS, is controlled by $\rho_0 / \rho_{\rm t}$ and leads to liquid-gas coexistence with binodal \textit{densities} $\rho_{g/\ell}$. Second, the instability driven by pairwise forces, PF-MIPS, is controlled by the packing fraction $\phi$ and the effective persistence length $\sim \ell_p / \rf$~\cite{Solon_2015_pressure_ABP,Omar_2023}, and promotes the coexistence between fixed \textit{packing fractions} $\phi_{g/\ell}$.} Since packing fraction $\phi$ and densities $\rho$ are related through $\phi=\rho \pi \rf^2/4$, the relative positions of these transitions evolve as $\rf$ varies,  leading to the rich physics summarized on the phase diagrams shown in Figs.~\ref{fig:phase_diag_densification}a-b. 

\begin{figure*}
    \centering
    \includegraphics[width=\textwidth]{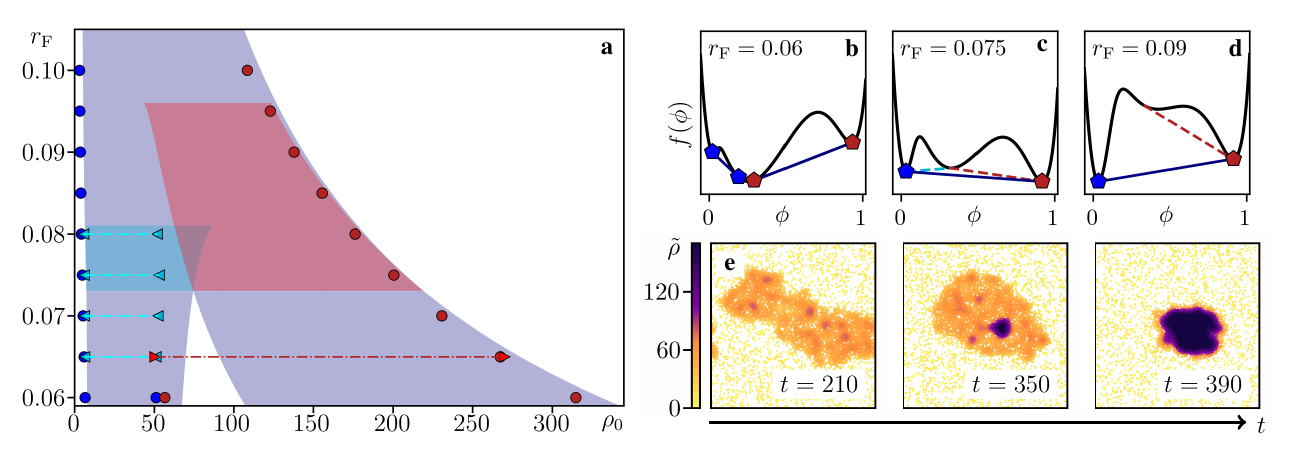}
    \caption{Metastable coexistence: comparison between theory and simulations. (a) Enlargement of the phase diagram of Fig.~\ref{fig:phase_diag_densification} for $\rf \in [0.06, 0.10]$. All symbols refer to numerical measurements of binodals. Circles represent stable binodals; left-pointing cyan triangles stand for QS-metastable binodals, while right-pointing red triangles correspond to PF-metastable binodals. Tie lines connect these coexisting metastable phases. Numerical results are compared to the analytical predictions for the binodals (shaded regions), obtained from Eq.~\eqref{eq:local_theory}: different colors represent stable (blue), QS-metastable (cyan) and PF-metastable (red) regions. (b-d) Common-tangent construction on the effective free energy $f(\phi)$ built from the local theory Eq.~\eqref{eq:local_theory}. Colored pentagons correspond to theoretical predictions of the free energy, and are connected by the supporting tangent to $f(\phi)$. Dashed tangent lines connect metastable binodals. (b) At low $\rf$, the free energy admits two pairs of stable binodals. (c-d) In the metastable region, either two pairs (c) or one pair (d) of metastable binodals can be observed. (e) Time-evolution of a simulation initialized in the QS-metastable region at $\rho_0=20,\;\rf=0.08$. Initially, QS induces MIPS between a pair of low-density binodals. Pairwise forces then promote the nucleation of a dense bubble within the QS-liquid region, eventually leading to a dramatic increase of the liquid density. 
    See End Matter and SM for numerical details~\cite{supp}. 
    }
    \label{fig:free_energy}
\end{figure*}

For small values of $\rf$, the phase diagram exhibits two  separate binodal regions: At small densities, QS-MIPS induces coexistence between a gas at $\rho=\rho^{\rm QS}_{g}$ and a liquid at $\rho=\rho^{\rm QS}_{\ell}$; At larger densities, PF-MIPS induces the coexistence between gas and liquid binodals denoted $\rho^{\rm PF}_{g}$ and $\rho^{\rm PF}_{\ell}$, respectively, as illustrated in the inset of Fig.~\ref{fig:phase_diag_densification}a. 
As $\rf$ increases, the quorum-sensing binodal densities (Fig.~\ref{fig:phase_diag_densification}a) and the pairwise-forces binodal packing fractions (Fig.~\ref{fig:phase_diag_densification}b) remain essentially constant. 
As a consequence, the pairwise-forces binodal densities decrease as $\rf^{-2}$. At  $\rf\approx 0.06$, the quorum-sensing liquid binodal and the pairwise-forces gas binodal meet and the two phase transitions merge. The dilute quorum-sensing phase at $\rho^{\rm QS}_{g}$ now coexists with the dense pairwise-force liquid at $\rho^{\rm PF}_{\ell}$. Remarkably, repulsive interactions thus densify dramatically the liquid phase of QS-MIPS, as opposed to what was observed in Fig.~\ref{fig:Fig2}a for $v_{\rm min}=0$. 
Let us now show how this qualitative understanding the phase diagram can be supported theoretically.

\bigskip
{\it Local hydrodynamic theory.}
To account for the observed phenomenology, we coarse-grain the dynamics of our system within a mean-field approximation~\cite{supp}. To describe the large-scale fluctuations of the density field, we resort to a gradient truncation at order $\mathcal{O}(\nabla^2)$. 
This amounts to  taking a local approximation for the self-propulsion speed: $v(\tilde \rho) \simeq v(\rho) + \mathcal{O}(\nabla^2)$. {We expect this approximation to work well when the correlation length is large enough for interfaces to be smooth, while the system remains sufficiently far from a critical point for a mean-field description to be quantitative~\cite{chaikin1995principles}. Very close to criticality, or far outside this regime, we expect qualitative agreement between numerical simulations and the local theory, but quantitative discrepancies~\cite{Solon_2018_Rmap}.

Under these approximations,} the dynamics of the density field reads $\partial_t \rho = -\partial_\alpha J_{\alpha}$, where the current $J_{\alpha}$ is driven by the gradient of an effective chemical potential $\mu^{\rm eff}(\rho)$:
\begin{align}\label{eq:local_theory}
          \!\! J_{\alpha}=-v\da 
     \mu^{\rm eff}\,, \quad \mu^{\rm eff} = \frac{\rho v^*(\rho)}{2D_r} +\!\! \int^{\rho}   \!\!\!  ds \, \frac{\partial_s p^{\rm IK}(s)}{v(s)}  \;.
\end{align}
Here we have introduced the effective speed $v^*(\rho) = \avgline{\dot \br_i \cdot \bu_i}$, which accounts for self-propulsion reduction due to both {quorum sensing} and collisions. 
In the absence of {pairwise forces}, $v^*(\rho)=v(\rho)$, while, in the absence of quorum sensing, it describes the standard slow-down of active particles induced by repulsive forces~\cite{Fily_2012_origin,Solon_2015_pressure_ABP}. Note that pairwise forces also enter the effective chemical potential $\mu^{\rm eff}$ through the direct pressure, defined as $p^{\rm IK} = -{\rm Tr}(\sigma^{\rm IK}_{\alpha \beta})/2$, where $\sigma^{\rm IK}_{\alpha \beta}$ is the Irving-Kirkwood stress tensor~\cite{irving1950statistical}.  

Equation~\eqref{eq:local_theory} thus accounts simultaneously for quorum sensing and pairwise forces. Direct algebra shows that  a homogeneous profile at density $\rho_0$ is linearly unstable when:
\begin{equation}\label{eq:linear_stability}
      \left. \frac{{\rm d} p^{\rm IK}}{{\rm d} \rho} + \frac{\rho v(\rho) v^*(\rho)}{2D_r} \left[ \frac{1}{\rho} + (\log v^*)' \right] \right|_{\rho=\rho_0} < 0  \;.
\end{equation}
We note that, in the absence of pairwise forces, Eq.~\eqref{eq:linear_stability} reduces to $\partial_{\rho}[\rho v(\rho)] <0 $, which is the standard instability condition for QS-MIPS~\cite{tailleur2008statistical,cates2015motility}.
In Fig.~\ref{fig:phase_diag_densification} we report the spinodals for both QS- and PF-MIPS predicted from Eq.~\eqref{eq:linear_stability}, which predicts the merging of the two phase transitions. The densification of the quorum-sensing liquid phase can thus be seen as resulting from a 
secondary MIPS instability induced  by pairwise forces. 

Note that Eq.~\eqref{eq:local_theory} also allows predicting the coexisting binodals. {Within our local theory,} $\mu^{\rm eff}$ depends solely on $\rho$, {which allows us to} introduce a generalized free energy $\mathcal{F}[\rho] = \int {\rm d} \br \, f(\rho)$ such that $f'(\rho) = \mu^{\rm eff}(\rho)$. {Consequently, } $\mathcal{F}$ plays the role of a Lyapunov functional for the local dynamics: $\partial_t \mathcal{F} = - \int {\rm d} \br \, v(\rho) (\nabla \mu^{\rm eff})^2 < 0$. The system thus evolves towards the minima of $\mathcal{F}$, and {the coexisting densities can be predicted using} a standard common-tangent construction~\cite{yeomans1992statistical,Solon_2018_Rmap,obyrne2023introduction}. 
{Neglecting the gradient terms in the local approximation is key to the existence of the free energy $\mathcal{F}$, but it is known to have a quantitative impact on the predicted binodals~\cite{wittkowski2014scalar,Solon_2018_Rmap,Speck_2021,Omar_2023}. Nevertheless, we observe in Fig.~\ref{fig:phase_diag_densification} a remarkable agreement between local theory and simulations. 
In SM~\cite{supp}, we show how, building on established methods~\cite{Solon_2018_Rmap,Omar_2023}, we can further improve this agreement by resorting to a higher-order gradient expansion.}

The analysis of the free energy also reveals that, when the two instabilities merge {at a fluid-fluid-fluid triple point}, the densification should be accompanied by metastability. Figure~\ref{fig:free_energy}a shows a close-up of the phase diagram near the transition region, where the local theory predicts QS-MIPS (cyan) and PF-MIPS (red) to be metastable. In Figs.~\ref{fig:free_energy}b-d, we show the evolution of the free-energy density $f(\phi)$ as $\rf$ increases.
For $\rf \leq 0.06$, the convex hull of the free-energy density $f(\phi)$ accommodates two separate common-tangent lines corresponding to stable QS- and PF-MIPS, see Fig.~\ref{fig:free_energy}b. 
As $\rf$ increases, however, the phase-coexistence scenarios become more complex, as illustrated by  Figs.~\ref{fig:free_energy}c-d. {First, the free-energy density $f(\rho)$ displays a common tangent line joining three coexisting densities, indicating the existence of a triple point. Upon further increasing $\rf$, $f(\rho)$} exhibits multiple common tangents, some of which lay strictly inside the convex hull of $f$, hence indicating metastable phase coexistence. 
Fig.~\ref{fig:free_energy}e and SM Movie~4 show that metastability is indeed observed numerically: starting from a homogeneous phase, the system first undergoes a stable QS-MIPS, until a dense droplet nucleates at long time in the liquid phase, leading to a further densification of the liquid phase. {Note that metastability does not rely on the existence of a free energy and could be accounted for outside our local approximation, e.g. using large-deviation theory as in~\cite{omar2021phase}.}

{Finally, we mention that metastability emerges because two phase transitions relying on distinct interactions---repulsive forces and motility regulation---merge. This makes this scenario robust and generic in systems where these interactions are present, as illustrated in End Matter for the case of chemotaxis. How passive systems can be engineered to induce a free energy density akin to that shown in Fig.~\ref{fig:free_energy} is an interesting open question in passive self-assembly. Whether the resulting dynamics would differ in active and passive systems due to non-equilibrium gradient terms is an interesting question that has recently attracted attention in simpler contexts~\cite{zakine2023minimum,cates2023classical}.}

\textit{Conclusions.} In this Letter we have shown that pairwise forces, which are typically neglected in the presence of mediated interactions, can lead to a rich and versatile physics. On the one hand, they can restore ergodicity in an arrested system by  promoting phase coexistence. On the other hand, they can have the opposite effect and induce a densification of an already phase-separated system by triggering a secondary instability in the liquid phase. 
{It would be interesting to see if our formalism could prove helpful to account for the reentrant phase transition recently reported in systems of self-propelled particles interacting via quorum sensing and pairwise forces~\cite{de2025quorum}.}

Accounting analytically for the emerging behaviors of nonequilibrium systems is  notoriously difficult and we have shown how to build a hydrodynamic description  that allows for predicting the phase transitions observed numerically. In Fig.~\ref{fig:phase_diag_densification}, we have shown how a local theory  accounts semiquantitatively for the observed phenomenology. In~\cite{supp}, we show how further improvement can be obtained by retaining higher-order gradients, in the spirit of~\cite{Solon_2018_Rmap,Speck_2021,Omar_2023}.

Beyond their theoretical interest, we believe that our results have implications for biological systems, where mediated interactions are ubiquitous. For instance,  bacteria like \textit{M. Xanthus} can undergo sporulation and form dense, quiescent aggregates~\cite{zusman2007chemosensory,liu2019self}. 
It is known that mediated interactions between bacteria are crucial to the formation of these aggregates, whose densities are very high. Furthermore, outside the biological realm, our results should be relevant for the design and control of synthetic active materials, where short-range and long-range interactions often compete~\cite{bricard2013emergence,nishiguchi2015mesoscopic,lefranc2025quorum}. For instance, density-based control of motility can be implemented via optical feedback loops~\cite{bechinger2016active,bauerle2018self}, which offers an interesting experimental framework to investigate the competition between quorum sensing and pairwise forces.

\section{Acknowledgments}
We thank Denis Bartolo for fruitful discussions. JT and QN thank laboratoire MSC for hospitality. 
QN is supported by the MIT Undergraduate Research Opportunities Program (UROP). AD acknowledges an international doctoral fellowship from Idex Université Paris Cité and a postdoctoral contract at the University of Geneva. GS acknowledges the support of the LabEx ‘Who Am I?’ and of the Universit\'e Paris Cit\'e IdEx funded by the French Government through its ‘Investments for the Future’ program. GS acknowledges support from the ERC Advanced Grant ActBio (funded as UKRI Frontier Research Grant EP/Y033981/1).

\textit{Data availability.} All numerical data are openly available at~\cite{PRLdata}.

~\clearpage{}

\section{End matter}
\textit{Simulation details.}
To integrate the dynamics corresponding to Eq.~(1), we use a Euler scheme with adaptive time steps. At each iteration, the time step is set to $\Delta t = \min \big [ \Delta t_{\rm max}, \frac{\rf}{10 v_{\rm M}} \big ]$, where $\Delta t_{\rm max}=10^{-4}$ in all our simulations,  
$v_{\rm M}$ is the magnitude of the largest particle velocity at time $t$, and $\rf$ is the range of the pairwise force interaction potential.
To compute the particle self-propulsion speed, we estimate the local density using a coarse-graining kernel $K$ given by$K(\vec{r})=\frac{1}{Z}\exp\left(\frac{r_{\rm QS}^2}{r_{\rm QS}^2-r^2}\right)\Theta(\rq-r)$ with $r_{\rm QS}=1$ and $Z$ a normalization constant such that $\int d\br K(\br)=1$. Finally, the soft repulsive potential takes the form
\begin{equation}
\label{eq:potential}
    V(\br) = \varepsilon \> \rf \left( 1-\frac{r}{\rf} \right)^2 \Theta(\rf - r)
\end{equation}
with $\varepsilon=100$. Equation~\eqref{eq:potential} ensures that the amplitude of the repulsive force is constant as we change $\rf$, hence allowing us to tune the range of the force without altering its scale.

\begin{figure}
    \centering
    \includegraphics[width=\columnwidth]{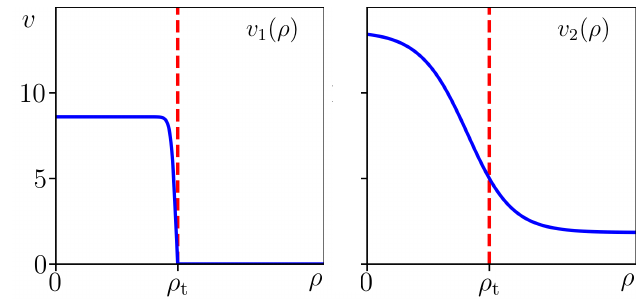}
    \caption{Schematic plots of $v_{1,2}(\rho)$. }
    \label{fig:speed_plots}
\end{figure}

To study the absorbing phase transition and its arrest, as in Figs.~\ref{fig:Fig1}a-b and Fig.~\ref{fig:Fig2}, the functional form of the self-propulsion speed $v(\rho)$ is chosen as:
\begin{equation}
    v_1(\rho) = v_0 \left\{ \exp\left[ -\lambda \tanh \left( \frac{\rho-\rho_{\rm t}}{\varphi} \right) \right] - 1 \right \} \Theta(\rho_{\rm t}-\rho) \;,
\end{equation}
which ensures $v=v_{\rm min} = 0$ beyond the threshold density $\rho_{\rm t}$. 
Finally, to study the repulsion-induced densification, as in  Figs.~\ref{fig:Fig1}c-d,~\ref{fig:phase_diag_densification} and~\ref{fig:free_energy}, we use:
\begin{equation}
    v_2(\rho) = v_0 \exp\left[ -\lambda \tanh \left( \frac{\rho-\rho_{\rm t}}{\varphi} \right) \right] \;.
\end{equation}
Both choices are illustrated on Fig.~\ref{fig:speed_plots}.

\textit{Figure parameters.}
For all figures, $D_r=1$, $\lambda =1$, $v_0=5$, and $\rho_{\rm t} = 25$.  
Other parameters are as follows: {
\begin{itemize}
    \item Fig.~\ref{fig:Fig1}:  $L_x=L_y=20$. Top panels: $v(\rho)=v_1(\rho),\varphi=1$, and $\rho_0=9$. Bottom panels: $v(\rho)=v_2(\rho), \varphi=10$, and $\rho_0=20$.
    \item Fig.~\ref{fig:Fig2}:  $L_x=L_y=40, v(\rho)=v_1(\rho)$, and $\varphi=1$. {The initial density is $\rho_0=9.6$ up to $\rf = 0.28$, where $\rho_0=9.6$ corresponds to a packing fraction $\phi_0\approx 0.65$. ($\rf = 0.28$ is above the transition at $\rf=0.15$.) 
    For larger $\rf$, $\rho_0$ is set such that $\phi_0\approx 0.65$ and the initial condition remains between the MIPS binodals.}
    \item Fig.~\ref{fig:phase_diag_densification}:  $L_x=L_y=20, v(\rho)=v_2(\rho)$, and $\varphi=10$. The system is initialized in an already phase-separated configuration with a liquid droplet of radius $R_{\rm d}=3$ coexisting with a homogeneous gas phase, and then allowed to relax. In the inset, $R_{\rm d}$ is adjusted so that $\rho_0=140$.
    \item Fig.~\ref{fig:free_energy}:  $L_x=L_y=20, v(\rho)=v_2(\rho)$, and $\varphi=10$. To explore the phase diagram, the system is initialized in an already phase-separated configuration with a liquid droplet of radius $R_{\rm d}=3$ coexisting with a homogeneous gas phase, and then allowed to relax. For panel (e), the system is initialized homogeneously at $\rho_0=20$.
\end{itemize}
}

\begin{figure}
    \centering
    \includegraphics[width=1.0\columnwidth]{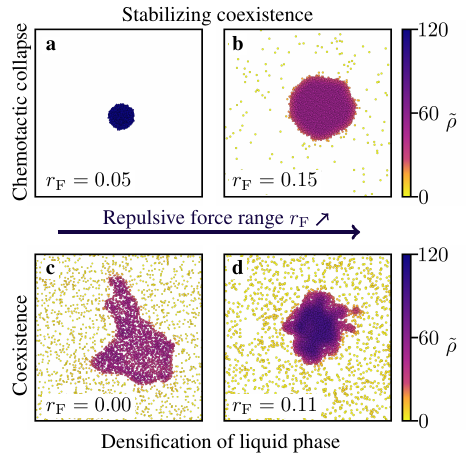}
    \caption{{Impact of repulsive forces of range $\rf$ on the self-organization induced by tactic interactions in a system of RTPs~\eqref{eq:taxis}.
    {\bf (a)} Chemotactic collapse in the presence of strong chemo-attractive interactions. Note that the liquid density saturates the colorbar.
    {\bf (b)}  Repulsive forces then oppose the arrest and stabilize a liquid-gas coexistence 
    {\bf (c)} Taxis-induced liquid-gas coexistence. 
    {\bf (d)} Repulsive forces induce a densification of the liquid phase.
    In panels (a, b) we use $c(\br)=c_1(\br)$, given in Eq.~\eqref{eq:c1}, $\alpha_0=0.2$, $\alpha_1=40$, $1/\ell_a=10$, $1/\ell_r=4$, $\epsilon_r/(2\pi)=0.2$, $\epsilon_a/(2\pi)=0.05$, $v_0=1$, and $\rho_0=6.3$. 
    In panels (c, d), we use $c(\br)=c_2(\br)$, given in Eq.~\eqref{eq:c2}, $\alpha_0=1$, $\alpha_1=-5$, $v_0=2$, $\rho_{\rm t}=25$, $\varphi=1$, $\rho_0=13$. In all panels, $L_x = L_y = 20$, and the repulsive potential $V$ is given by Eq.~\eqref{eq:potential} with $\varepsilon=100$.}}
    \label{fig:taxis}
\end{figure}

{
\emph{Interplay between pairwise forces and chemotaxis}. 
Here we show that the phenomenology of Fig.~\ref{fig:Fig1} is also observed for run-and-tumble particles (RTPs) with chemotactic interactions.
We consider RTPs whose tumbling rates are biased by the gradients of a chemical field $c$~\cite{budrene1991complex,schnitzer1993theory,berg2004coli,o2020lamellar}, so that their dynamics read:
\begin{subequations}
    \label{eq:taxis}
    \begin{align}
    \label{eq:RTPdynamics}
    \dot{\vec{r}}_i &= v_0 \vec{u}_i(\theta_i)- \sum_j \vec{\nabla} V(\vec{r}_{i}-\vec r_{j})\;, \\
    {\theta}_i &\overset{\alpha}{\to} \theta'_i \quad \text{at rate} \quad \alpha = \alpha_0 + \alpha_1 \bu \cdot \nabla c\;,
    \label{eq:tumbling}
    \end{align}
\end{subequations}
where $v_0, \alpha_0, \alpha_1$ are constant and $V$ is given by Eq.~\eqref{eq:potential}.
In turn, the chemical field $c(\br)$ is produced and consumed by the particles. 
Assuming that $c$ diffuses rapidly, an adiabatic approximation allows us to express it as a function of the particle density field: $c = c(\br, [\rho])$. See, for instance, Ref.~\cite{o2020lamellar} for an explicit derivation.

In Figs.~\ref{fig:taxis}a-b, we study the impact of pairwise forces on a chemotactic collapse~\cite{brenner1998physical,o2020lamellar} that results from the competition between a strong chemoattractant and a weaker chemorepellent. 
This generalizes the results of Figs.~\ref{fig:Fig1}a-b for quorum sensing interactions.
Following Ref.~\cite{o2020lamellar}, we model the chemical field $c$ as:
\begin{equation}
    c_1(\br, [\rho]) = \epsilon_r G_{r} \ast \rho - \epsilon_a G_a \ast \rho ,
    \label{eq:c1}
\end{equation}
where the indices $a, r$ denote, respectively, chemo-attractant and chemo-repulsive signals, and the function $G_{a,r}(\br-\br')$ relates the bacterial density at position $\br'$ to the chemical fields at position $\br$. We use $G_n(\bfr) = (2\pi)^{-1} K_0(|\bfr|/\ell_n)$  where $K_0$ is the 0th order modified Bessel function of the second kind. 
Note that $G_n$ is the Green's function associated with the screened Poisson equation in two dimensions, so this choice models the steady-state behavior of a diffusive chemical signal with constant production and degradation rates.
The chemotactic collapse observed at small $\rf$ in Fig.~\ref{fig:taxis}a is stabilized into a liquid-gas coexistence upon increasing $\rf$, see Fig.~\ref{fig:taxis}b.

We then turn to the case of taxis-induced phase coexistence and its densification via pairwise forces. We study this case by modeling the chemical field through:
\begin{equation}
    c_2(\br, [\rho]) = \tanh\bigl[ \frac{\tilde\rho - \rho_{\rm t}}{\varphi}\bigr] \;,
    \label{eq:c2}
\end{equation}
where the definition of $\tilde\rho$ is the same as in the main text.
Here, Eq.~\eqref{eq:c2} describes a threshold response to the chemical signals produced by other particles, whose concentration is proportional to the local density.
Akin to the phenomenology observed in Fig.~\ref{fig:Fig1}c, this system undergoes phase separation in the absence of pairwise forces between finite-density gas and liquid phases, see Fig.~\ref{fig:taxis}c. Upon increasing the interaction range of repulsive contact forces, we observe a densification of the liquid phase in Fig.~\ref{fig:taxis}d, hence recovering the phenomenology of Fig.~\ref{fig:Fig1}d.
}

\onecolumngrid
\newpage
\begin{center}
  {\large\bfseries Supplemental Material for\\``Contact Forces in Motility-Regulated Active Matter"\par}
  \vspace{1em}
\end{center}

\renewcommand{\thesection}{S\Roman{section}}
\renewcommand{\theequation}{S.\arabic{equation}}
\renewcommand{\thefigure}{S\arabic{figure}}

\def\ito{It\^{o} }
\def \pfap/{PFAPs}
\def \qsap/{QSAPs}
\def \assume/{\textbf{ASSUME}}
\newcommand{\id}{\mathbbm{1}}

\renewcommand{\vec}[1]{\mathbf{#1}}
\renewcommand{\tensor}[1]{\vec{\vec{#1}}}
\newcommand{\tensorr}[1]{\vec{\tensor{#1}}}
\newcommand{\alphaa}{\vec{\tensorr{\alpha}}}
\newcommand{\dg}{\partial_{\gamma}}

\renewcommand{\rq}{r_{\rm QS}}
\newcommand{\pIK}{p^{\rm IK}}
\newcommand{\mueff}{\mu^{\rm eff}}

\let\addcontentsline\origaddcontentsline
\setcounter{secnumdepth}{3}

\setcounter{section}{0}
\setcounter{equation}{0}
\setcounter{figure}{0}
\setcounter{table}{0}
\setcounter{footnote}{0}
\renewcommand{\theHsection}{S\Roman{section}}
\renewcommand{\theHequation}{S.\arabic{equation}}
\renewcommand{\theHfigure}{S\arabic{figure}}
\renewcommand{\theHtable}{S\arabic{table}}

\tableofcontents
\bigskip
The Supplemental Material (SM) contains details on the derivation of the theoretical and numerical results presented in the main text. Throughout the SM, Eq.~(X) and Fig.~SX refer to equations and figures of the main text while Eq.~(S.X) and
Fig.~X refer to equations and figures that appear in the Supplemental Material.

In Sec.~\ref{sec:coarsegraining} we detail the coarse-graining procedure of the microscopic dynamics to obtain the large-scale field theory. We then provide a closure to the hydrodynamic equations using the leading-order gradient truncation, as described in Sec.~\ref{sec:local}. The resulting \textit{bona-fide} equilibrium theory, corresponding to Eq.~(3), is used in Sec.~\ref{sec:phase_coex} to predict the binodals corresponding to Figs.~2-3. 
In Sec.~\ref{SMsec:grad_theory} we then show how these predictions can be further improved by deriving a higher-order gradient theory. The resulting field theory being out-of-equilibrium, in Sec.~\ref{sec:Rmapping} we follow~\cite{SM-Solon_2018_Rmap,SM-Omar_2023} to predict the coexisting densities. Next, in Sec.~\ref{sec:absorbing} we move to the absorbing phase transition presented in Sec.~``Repulsive forces oppose condensation and arrest." of the main text. 
We show the robustness of the kinetic argument for phase separation by testing it over different sets of parameters in Sec.~\ref{sec:test}. In Sec.~\ref{sec:refined} we refine the kinetic argument by taking into account the growth dynamics of the droplet. We conclude this section by providing details in Sec.~\ref{sec:MIPS} on how to derive the binodals in the regime where pairwise forces dominate over quorum sensing, which corresponds to the orange-shaded region of Fig.~2. Finally, in Sec.~\ref{SMsec:movies} we report all captions and parameters for the Supplementary Movies.

\newpage

\section{From the microscopic dynamics to the local hydrodynamic theory}
In this section we coarse grain the microscopic dynamics of active Brownian particles interacting via quorum sensing and pairwise forces, described by Eq.~(1) of the main text. We then derive the corresponding mean-field, local hydrodynamic theory given in Eq.~(3).  Starting from this equation, we then predict the gas and liquid binodals. 

\subsection{Coarse-graining procedure}
\label{sec:coarsegraining}
Our starting point is the microscopic dynamics of ABPs with quorum-sensing and repulsive interactions, which we reproduce here for clarity. Denoting $\br_i$ and $\bu_i \equiv (\cos\theta_i, \sin\theta_i)$ the position and orientation of particle $i$, respectively, their dynamics reads:
\begin{equation}
\label{eq-app:dynamics}
    \dot{\vec{r}}_i = v[\tilde\rho(\br_i)] \bu_i - \mu_0 \sum_j \vec{\nabla} V(\vec{r}_{i}-\vec r_{j})\;, \quad
\dot{\theta}_i = \sqrt{2D_r} \eta_i(t)\;,
\end{equation}
where $\eta_i(t)$ is a centered Gaussian white noise with delta-correlations: $\langle \eta_i(t) \eta_j(t')\rangle = \delta_{ij}\delta(t-t')$. As detailed in the End Matter, the self-propulsion speed $v$ is a function of the effective density $\tilde\rho_i = (K \ast \rho)(\br_i)$, where $\ast$ denotes the convolution product and $K(r)$ is a normalized isotropic bell-shaped kernel. Finally, the unit of time are such that the particle mobility is set to $\mu_0 = 1$.

Our goal is to obtain a closed expression for the particle current $\bfJ(\br,[\rho])$ as a function of the particle density field $\rho(\br)$. To do so, we first introduce the fluctuating microscopic density 
\begin{equation}
\hat \psi(\br, \theta) = \sum\limits_i^N \delta(\br - \br_i) \delta( \theta - \theta_i)\;.
\end{equation}
Successive moments of $\hat{\psi}$ can be used to define the fluctuating density, polar and nematic fields as:
\begin{equation}
    \hat \rho(\br) = \int \rmd \theta \hat\psi(\br,\theta)\;,\qquad \hat m_{\alpha}(\br) = \int \rmd \theta u_{\alpha}\hat\psi(\br,\theta) \;, \qquad  \hat Q_{\alpha\beta}(\br) = \int \rmd \theta \biggl( u_{\alpha}u_{\beta} - \frac{\delta_{\alpha\beta}}{2}\biggr)\hat\psi(\br,\theta) \;.
\end{equation}
Throughout our discussion we use the hat notation, $\hat f$, for fluctuating quantities, while $f = \langle \hat f \rangle$ represents their average over noise.

Using \ito  calculus~\cite{SM-dean1996langevin}, the dynamics of $\hat \psi$ reads:
\begin{equation}
    \partial_t \hat{\psi} = - \nabla_{\vec{r}} \cdot \biggl[
     \hat{v}(\br) \vec{u}(\theta) \hat\psi  -\int \rmd \br' \nabla_{\vec{r}}V(\vec{r}-\vec{r}') \hat{\psi}(\br) \hat \rho(\br') \biggr]  + D_r \nabla_{\theta}^2\hat{\psi} - \nabla_{\theta}\biggl[ \sqrt{2D_r\hat{\psi} } {\Lambda_\theta} \biggr]\;,
    \label{eq:fluct_dynamics}
\end{equation}
where $\hat v(\br) \equiv v[\tilde \rho(\br)]$ and $\Lambda_\theta$ is a Gaussian white noise field satisfying:
\begin{equation}
\avg{\Lambda_\theta} = 0 \;, \quad \avg{\Lambda_\theta(\br,\theta,t) \Lambda_\theta(\br',\theta',t')} = \delta(\br-\br') \delta(\theta-\theta') \delta(t-t')  \;.  
\end{equation}
To get a closed expression for the dynamics of the average density field $\rho = \avg{\hat \rho}$, we integrate Eq.~\eqref{eq:fluct_dynamics} over $\theta$ and average over noise. We obtain:
\begin{equation}
    \dt \rho = -\da {J}_{\alpha} \;, \quad  J_{\alpha} \equiv \langle \hat J_{\alpha} \rangle = \db {\sigma}^{\rm IK}_{\alpha\beta} + \avg{ {\hat v}{\hat m}_{\alpha} } \;, \label{eq:rho_dyn}
\end{equation}
where $\sigma^{\rm IK}_{\alpha\beta}$ is the Irving-Kirkwood stress tensor~\cite{SM-irving1950statistical}, accounting for particle fluxes induced by steric forces, given by
\begin{equation}\label{app-eq:ik-stress}
   \tensor{\sigma}^{\rm IK}_{\alpha \beta} (\br) \equiv \langle \hat{\sigma}^{\rm IK}_{\alpha \beta} (\br) \rangle =  \avg{-\frac{1}{2}\sum_{ij} (\vec{r}_{ij})_{\alpha} \vec{F}_{\beta}(\vec{r}_{ij})  \int_0^1 \rmd s \, \delta(\vec{r} - \vec{r}_j - s(\vec{r}_i-\vec{r}_j)) }\;.
\end{equation}

We note that, at this stage, the hydrodynamic Eq.~\eqref{eq:rho_dyn} is not closed, in particular due to the correlation function involving density-dependent self-propulsion speed and polar field, $\langle \hat v {\hat m}_{\alpha} \rangle$.  
To provide a suitable closure,  we first derive the dynamics of $\hat m_{\alpha}$ by multiplying Eq.~\eqref{eq:fluct_dynamics} by $u_{\alpha}$ and then integrating over $\theta$. The dynamics of $\hat m_{\alpha}$ then reads:
\begin{equation}
    \dt \hat{m}_{\alpha} = -\db\left[\hat{\rm I}^{(m)}_{\alpha\beta} + \hat v (\hat{Q}_{\alpha\beta} + \hat\rho \delta_{\alpha \beta} /2)  \right] -D_r \hat{m}_{\alpha} + \Lambda_{m,\alpha}(\br,t) \;, \label{eq:m_dyn}
\end{equation}
where $\mathbf{\Lambda}_m(\br,t)$ is a zero-mean Gaussian field with covariance $\avg{\Lambda_{m,\alpha}(\br,t)\Lambda_{m,\beta}(\br',t')}=2D_r (\rho\delta_{\alpha\beta}/2-Q_{\alpha\beta}) \delta(\br-\br')\delta(t-t')$, and
\begin{equation}
    \hat{\rm I}^{(m)}_{\alpha \beta}(\br) =
    \int \rmd \br' F_{\beta}(\br-\br') \hat m_{\alpha}(\br) \hat \rho(\br')\;. 
    %
\end{equation}
We now evaluate the correlation term $\avg{\hat v \hat{m}_{\alpha}}$ entering Eq.~\eqref{eq:rho_dyn}. To do so, we first use \ito calculus to derive its dynamics:
\begin{equation}\label{eq:cg-step_corr}
    \dt \avg{\hat v \hat{m}_{\alpha}} = -D_r \avg{\hat v \hat{m}_{\alpha}} - \avg{ \hat v \db \left[ \hat v \left(\frac{1}{2}\hat\rho\delta_{\alpha \beta}+ \hat{Q}_{\alpha \beta}\right)+\hat {\rm I}^{(m)}_{\alpha\beta} \right]} - \avg{\hat{m}_{\alpha} ~ \hat{v}' K \ast \db \hat J_{\beta} }\;, 
\end{equation}
where the last term results from $\partial_t \hat v = \partial_t  v(K \ast \hat\rho) = \hat v' K \ast \partial_t \hat\rho = - \hat v' K \ast \partial_\alpha \hat J_\alpha$. Equation~\eqref{eq:cg-step_corr} shows how the term $\avg{\hat v \hat{m}_{\alpha}}$ has a finite decay time $1/D_r$. Meanwhile, the density field is a hydrodynamic mode and, as such, its fluctuations relax over a timescale $T \sim L^\gamma$, where $L$ is the linear system size and $\gamma$ a positive exponent. Using a fast variable approximation over $\avg{\hat v \hat{m}_{\alpha}}$ then turns Eq.~\eqref{eq:cg-step_corr} into
\begin{align}
    \avg{\hat v \hat{m}_{\alpha}} = - \frac{1}{D_r}\avg{ \hat v \db \left[ \hat v \left(\frac{1}{2}\hat\rho\delta_{\alpha \beta}+ \hat{Q}_{\alpha \beta}\right)+\hat {\rm I}^{(m)}_{\alpha\beta} \right]} - \frac{1}{D_r}\avg{\hat{m}_{\alpha} \hat{v}' ~ K * \db \hat J_{\beta}}\;. \label{eq:full_gradients}
\end{align}
Inserting Eq.~\eqref{eq:full_gradients} inside Eq.~\eqref{eq:rho_dyn} then yields:
\begin{equation}
    \dt \rho = -\da \left\{ \db \sigma^{\rm IK}_{\alpha\beta} - \frac{1}{D_r}\avg{ \hat v \db \left[ \hat v \left(\frac{1}{2}\hat\rho\delta_{\alpha \beta}+ \hat{Q}_{\alpha \beta}\right)+\hat {\rm I}^{(m)}_{\alpha\beta} \right]} - \frac{1}{D_r}\avg{\hat{m}_{\alpha} \hat{v}' ~ K * \db \hat J_{\beta}} \right\} \;.
    \label{eq:full_hydrodynamics}
\end{equation}
To rewrite Eq.~\eqref{eq:full_hydrodynamics} in a more manageable form, we first introduce the generalized active stress tensor:
\begin{equation}
    \sigma^{\rm a}_{\alpha\beta} \equiv \langle \hat\sigma^{\rm a}_{\alpha\beta} \rangle = -\frac{1}{D_r}\avg{ \hat v^2 \left(\frac{1}{2}\hat\rho\delta_{\alpha \beta}+ \hat{Q}_{\alpha \beta}\right)+ \hat v \hat {\rm I}^{(m)}_{\alpha\beta} }\;.
\end{equation}
We note that $\hat\sigma^{\rm a}_{\alpha\beta}$ has the same functional form as the standard active stress tensor for active particles interacting only via pairwise-forces~\cite{SM-solon2015pressure,SM-Speck_2021,SM-Omar_2023}, albeit with $\hat v$ replacing a constant speed $v_0$. Finally, defining the generalized stress tensor $\sigma_{\alpha \beta} = \sigma^{\rm IK}_{\alpha \beta} + \sigma^{\rm a}_{\alpha \beta} $, we rewrite Eq.~\eqref{eq:full_hydrodynamics} as:
\begin{equation}
    \dt \rho = -\da J_{\alpha}\;, \qquad J_\alpha = \db \sigma_{\alpha\beta} - \langle \hat{\sigma}^a_{\alpha\beta} \partial_\beta \log \hat{v} \rangle  - \frac{1}{D_r}\avg{\hat{m}_{\alpha} \hat v'~K \ast \db \hat{J}_{\beta}} \;.
    \label{eq:sigma_hydro}
\end{equation}
Eq.~\eqref{eq:sigma_hydro} can now be used to derive the mean-field `local' hydrodynamics, where `local' refers to the fact that we work at the leading order in a gradient expansion. The role of higher-order gradients is then discussed in Sec.~\ref{SMsec:grad_theory}.

\if{\subsection{Lack of an equation of state}
Before moving towards the prediction of the phase coexistence, we briefly comment on the lack of an equation of state for the dynamics~(1) and show how we can recover the purely pairwise-forces and quorum-sensing limit. \gs{Consider if having this section or not, we can copy paste what we wrote originally in the manuscript which was a nice discussion}}\fi

\subsection{Local hydrodynamic theory}\label{sec:local}
To derive a closed hydrodynamic theory, we first use a mean-field approximation of Eq.~\eqref{eq:sigma_hydro}, within which we neglect fluctuations and approximate the average of a function of a stochastic variable as $\langle G(\hat f) \rangle \approx G (f)$. 
Furthermore, we rely on gradient truncation to express $J_{\alpha}$ up to $\cO(\nabla)$ terms, as is customary to describe long-wavelength hydrodynamic modes. 
Note that these approximations are expected to hold in bulk phases, far from interfaces~\cite{SM-Solon_2018_Rmap}. 

We now detail how different terms in Eq.~\eqref{eq:sigma_hydro} simplify within our approximations:
\begin{itemize}
    \item Locality of the interactions allows us to write $v(\tilde \rho) \approx v(\rho) + \cO(\nabla_\br^2)$.
    \item Since the system is scalar and isotropic, we express $\sigma_{\alpha \beta}^{\rm IK} = - p^{\rm IK}(\rho) \delta_{\alpha \beta} + \cO(\nabla_\br)$ where $p^{\rm IK}(\rho) = - \sigma^{\rm IK}_{\gamma \gamma}/2$ is the Irving-Kirkwood pressure. 
    \item We now consider the generalized active stress tensor. We first rewrite it as:
\begin{align*}
    \sigma^{\rm a}_{\alpha\beta} &= \avg{ -\frac{\hat v[\tilde \rho]}{D_r} \left[ \hat v[\tilde \rho] \sum_i \delta(\br-\br_i) u_{i,\alpha}u_{i,\beta} + \sum_i \delta(\br - \br_i)u_{i,\alpha}\sum_j F_{\beta}(\br_{ij}) \right]} = \avg{-\frac{\hat v[\tilde \rho]}{D_r}\sum_i \delta(\br-\br_i)u_{i,\alpha}\dot r_{i,\beta}}\;,
\end{align*}
where the sum runs over particles $i$. At mean-field level, we factorize the product inside the average as:
\[
\sigma^{\rm a}_{\alpha\beta} \approx -\frac{v(\rho)}{D_r} \avg{\sum_i \delta(\br-\br_i)u_{i,\alpha}\dot r_{i,\beta}}\;.
\]
Finally, for a homogeneous isotropic system or, more generically, for bulk phases far from any interface, $\sigma^{\rm a}_{\alpha\beta}$ simplifies into:
\begin{align}
\sigma^{\rm a}_{\alpha\beta} &= -p^a \delta_{\alpha\beta}\qquad\text{with}\qquad  p_a = -\frac{v(\rho)}{D_r} \rho \avg{\dot\br \cdot \mathbf{u}} \delta_{\alpha\beta} + \cO(\nabla^2) \equiv -\frac{\rho v(\rho) v^*(\rho)}{2D_r}\delta_{\alpha\beta} \;,
\label{eq:homogeneous_approx} 
\end{align}
where we have introduced the generalized active pressure $p^a$ and the reduced effective speed of the particles in the bulk $v^*(\rho) \equiv \avg{\dot{\vec{r}}\cdot\vec{u}}$. In the absence of quorum sensing, the effective speed is well described by $v_{\rm PF}^*(\rho) \equiv v_0 U(\rho)$, where $U(\rho)$ is a dimensionless function accounting for the reduction in the self-propulsion speed induced by repulsive interactions~\cite{SM-stenhammar2013continuum,SM-Solon_2015_pressure_ABP,SM-arnoulx2019active,SM-Speck_2021}. 
We then generalize this expression to QS active particles as:
\begin{equation}
     v^*(\rho) = v(\rho)U(\rho) \;, \quad \text{so that} \quad \sigma^{\rm a}_{\alpha\beta} = -\frac{\rho v(\rho)^2}{2 D_r} U(\rho) \delta_{\alpha\beta} + \cO(\nabla_\br)\;.
\end{equation}
    \item Finally, the last term in Eq.~\eqref{eq:sigma_hydro} is negligible within our approximations, since:
    \begin{equation}
        \avg{\hat{m}_{\alpha} \hat v'~K \ast \db \hat{J}_{\beta}} \approx {m}_{\alpha} v'~K \ast \db {J}_{\beta} = \cO(\nabla_\br^2) \;.
    \end{equation}
\end{itemize}

All in all, to leading order in gradients, the hydrodynamics Eq.~\eqref{eq:sigma_hydro} reduces to:
\begin{align}
    \dt \rho = 
    \da J_{\alpha} \;, \quad J_{\alpha} =  \da p^{\rm IK} + v(\rho) \da\left[\frac{\rho v^*(\rho)}{2D_r} \right]   + \cO(\nabla^2)  \;. \label{eq:1st_local_theory}
\end{align}

Finally, we factor $v(\rho)$ and rewrite Eq.~\eqref{eq:1st_local_theory} to get Eq.~(3) of the main text as
\begin{equation}
    \dt \rho = \da [ v(\rho) \da \mu^{\rm eff}] \;, \qquad \mu^{\rm eff}(\rho) = \frac{\rho v^*(\rho)}{2D_r}  + \int^{\rho} ds \frac{\partial_s p^{\rm IK}(s)}{v(s)}\;.\label{eq:1st_local_binodal}
\end{equation}

\subsection{Predicting the binodals from the local theory}
\label{sec:phase_coex}
In this section we show how to predict the coexisting binodals, as in Fig.~3 of the main text, from the local theory~\eqref{eq:1st_local_binodal}. Since Eq.~\eqref{eq:1st_local_binodal} is an equilibrium theory, we can extract the free energy density $f(\rho)$ from the effective chemical potential $\mu^{\rm eff}$ and apply the equilibrium common-tangent construction to determine the binodals of phase coexistence~\cite{SM-chaikin1995principles}.

To do so, we need a closed expression for $\mu^{\rm eff}(\rho)$ as the density $\rho$ and the pairwise-force interaction range $\rf$ are varied. Consequently, we need  a functional form for both the direct pressure $p^{\rm IK}(\rho,\rf)$ and the reduced effective self-propulsion speed $v^*(\rho,\rf) = v(\rho)U(\rho, \rf)$. We assume $U$ to be a function of the packing fraction $\phi=\frac{\pi}{4}\rho\rf^2$ only, as this term encodes the slowing down due to inter-particle collisions. We thus take the following expression for the effective self-propulsion speed:
\begin{equation}
    v^*(\rho,\rf)=v(\rho) U(\phi)\;,
    \label{eq:pact_fit}
\end{equation}
where $U(\phi)$ is specified later. 

For the Irving-Kirkwood pressure $p^{\rm IK}$, we start from the case of pairwise-forces active particles (PFAPs), whose pressure has been extensively studied in the literature~\cite{SM-Solon_2015_pressure_ABP,SM-Solon_2018_Rmap,SM-Speck_2021,SM-Omar_2023}, and extend existing theory to the presence of quorum-sensing interactions. Standard scaling arguments~\cite{SM-Solon_2015_pressure_ABP,SM-martin2021statistical} show that one can express the Irving-Kirkwood pressure for a system of PFAPs with constant self-propulsion $v$ as: $p^{\rm IK, PF} = \frac{v}{\rf} h(\phi)$, where $h$ is a dimensionless scaling function~\footnote{$h$ is dimensionless because we set $\mu_0=1$.}. In the presence of quorum sensing we expect the Irving-Kirkwood pressure to scale with the local self-propulsion speed $v(\rho)$, so that we formulate the following ansatz:
\begin{equation}
    p^{\rm IK}(\rho,\rf) = 
    \frac{v(\rho)}{\rf} h(\phi) \;.
    \label{eq:pIK_fit}
\end{equation}

We now perform microscopic simulations to measure both $v^*$ and $p^{\rm IK}$ in our system, following Ref.~\cite{SM-Zhao_2024_Young_Dupre}.
We consider homogeneous systems of size $L_x=L_y=10$, and vary $\rf$, $\lambda,$ $\rho$, and $\ell_p^{\rm eff} \equiv v_1(\rho=0)/\rf D_r$. 
(See End Matter for the functional form of $v_1$.) We then measure the effective self-propulsion speed $v^*(\rho) = \langle \dot\br_i  \cdot \bu_i\rangle_{\{i,t\}}$, where the average is performed over particles $i$ and time $t$. Next, to determine $p^{\rm IK}$ we measure the IK stress tensor Eq.~\eqref{app-eq:ik-stress} on a grid of square cells indexed by $(n,m)$ using:
\begin{equation}
    \sigma^{\rm IK}_{\alpha\beta}(n,m)=\frac{1}{2r_0^2}\sum_{i \vee j}\tilde{r}_{ij,\alpha}\partial_\beta V(\br_i-\br_j) \;,\label{eq:measIK}
\end{equation}
where $r_0=1$ is the side length of a cell, $\tilde{\br}_{ij}$ is the fraction of $\br_{ij}$ that lies inside the $(n,m)$-th cell, and $i \vee j$ refers to pairs of particles such that the line segment going from $\br_i$ to $\br_j$ has a non-zero intersection with the cell $(n,m)$. The Irving-Kirkwood pressure is then given by $p^{\rm IK}(\rho) = \langle - \sigma_{xx}^{\rm IK} (n,m) \rangle_{\{(n,m),t\}}$.

\begin{figure}[h]
    \centering
    \begin{tikzpicture}
    \def\x{200pt}
        \path (0,0) node {\includegraphics[width=1.0\textwidth]{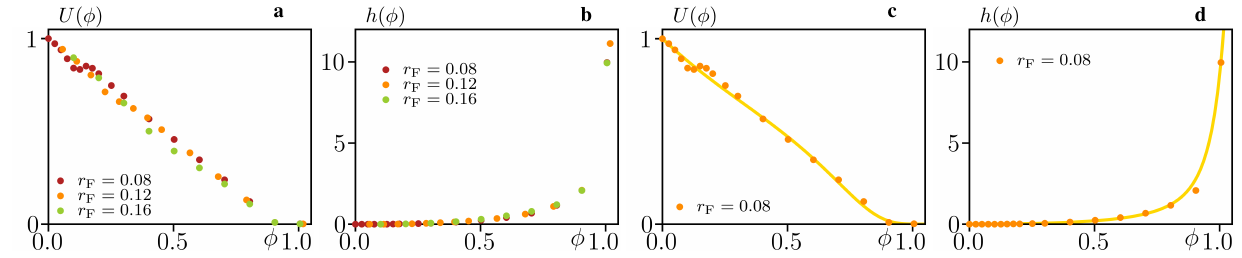}};
    \end{tikzpicture}
    \caption{({\bf a-b}) Numerical measurements of $U(\phi)$ and $h(\phi)$ with different values of the pairwise interaction range $\rf$, quorum-sensing parameter $\lambda$, and rotational diffusion $D_r$. The sets of parameters for each measurements are: (1) $\rf=0.08,\lambda=0.4,\ell_p^{\rm eff}=12.43$; (2) $\rf=0.12,\lambda=0.3,\ell_p^{\rm eff}=11.27$; (3) $\rf=0.16,\lambda=0.2,\ell_p^{\rm eff}=10.10$, where $\ell_p^{\rm eff}=v(\rho=0)/\rf D_r$. Measurements show how the two contributions collapse upon proper rescaling. {\bf (c-d)} Fits of $U(\phi)$ and $h(\phi)$ measured in simulations (orange circles) to Eq.~\eqref{eq:pressure-eq} using the PySR symbolic regression method, for $\rf=0.08$.
    }
    \label{fig:pressure_measurements}
\end{figure}

In Figs.~\ref{fig:pressure_measurements}a-b we show how numerical measurements of $U(\phi)$ and $h(\phi)$ collapse upon proper rescaling of the parameters, hence supporting our choices of the scalings for $v^*$ and $p^{\rm IK}$. Due to the lack of analytical expressions for $U,h$, we fit numerical data to Eqs.~\eqref{eq:pact_fit},~\eqref{eq:pIK_fit} using the Python package PySR that performs symbolic regression. This allows us to find an expression for the functions $U, h$ with physical constraints. 
 For $U(\phi)$, we enforce $U(0)=1$ and $U(\infty)=0$, and only allow compositions and linear combinations of $e^{-x}$, $1-\tanh(x)$, $x^2$, and $x^3$. For $h(\phi)$, we enforce $h(0)=0$ and only allow compositions and linear combinations of $e^x-1$, $x^2$, and $x^3$. Fig.~\ref{fig:pressure_measurements} shows the best results that PySR produce after $10$ trials, yielding the following expressions:
\begin{equation}
    U(\phi) = \exp\left[ -\frac{4\phi/\pi}{1-\tanh(A\phi^3)} \right] \;, \qquad h(\phi) = D [\exp(B\phi^6)-1+\exp(C\phi^2)-1] \;,
    \label{eq:pressure-eq}
\end{equation}
with $A = 1.21, B = 2.75, C = 1.16, D=0.544$. With these expressions for $U$ and $h$, we now have a closed form for $\mu^{\rm eff}(\rho)$ from which we can extract the free energy density $f(\rho)$ and predict the binodals shown in Fig.~3. 

\section{Beyond the local approximation: higher-order gradient theory}\label{SMsec:grad_theory}

In this section, we discuss how to improve the predictions of the local theory by retaining the next-order terms in the gradient expansion leading to the hydrodynamic description. To do so, we start from Eq.~\eqref{eq:sigma_hydro} and extend the method introduced by Omar and co-workers in~\cite{SM-Omar_2023} to our system, where both quorum sensing and pairwise forces are present. Within the mean-field approximation, we work at steady state ($\partial_\alpha J_\alpha=0$) and study how the conditions for phase coexistence are affected by higher-order gradient terms. 

First, within a mean-field approximation, the last term in Eq.~\eqref{eq:sigma_hydro} vanishes at all orders in gradients when $\partial_\alpha J_\alpha=0$, since:
\begin{equation}
    \avg{\hat{m}_{\alpha} \hat v'~K \ast \db \hat{J}_{\beta}} \approx \avg{\hat{m}_{\alpha} \hat v'} K \ast \db {J}_{\beta} =  0 \;.
\end{equation}
Consequently, Eq.~\eqref{eq:sigma_hydro} becomes:
\begin{equation}
0=\db \sigma^{\rm IK}_{\alpha\beta} +
    \avg{\hat v \hat{m}_{\alpha}} \;, \qquad \text{with}\qquad \avg{\hat v \hat{m}_{\alpha}}= - \frac{1}{D_r}\avg{ \hat v \db \left[ \hat v \left(\frac{1}{2}\hat\rho\delta_{\alpha \beta}+ \hat{Q}_{\alpha \beta}\right)+\hat {\rm I}^{(m)}_{\alpha\beta} \right]}\;.
\label{eq:current_balance}
\end{equation}
Our goal is to provide a closed expression, at higher-order in gradients, for Eq.~\ref{eq:current_balance}.
Following~\cite{SM-Omar_2023}, we decompose $\hat{\rm I}^{(m)}_{\alpha \beta}(\br)= \sum_{ij}{F}_{\alpha}(\vec{r}_{ij}){u}_{i,\beta} \delta(\vec{r}-\vec{r}_i)$ in a body-force-like and a stress-like term:
\begin{subequations}
\label{eq-app:ansatz-m}
\begin{gather}
    \hat{\rm I}^{(m)}_{\alpha \beta} = \hat{\rm \kappa}^{(m)}_{\alpha\beta} + \partial_{\gamma}\hat\Sigma^{(m)}_{\alpha\beta\gamma}\;, \\[0.2cm]
    \hat{\rm \kappa}^{(m)}_{\alpha\beta}(\br) = \frac{1}{2}\sum_{ij}F_{\alpha}(\vec{r}_{ij}) ({u}_{i,\beta}-{u}_{j,\beta}) \delta(\vec{r}-\vec{r}_i)\;, \\[0.2cm]
    \hat\Sigma^{(m)}_{\alpha\beta\gamma} = -\frac{1}{2}\sum_{ij} {r}_{ij,\gamma} {F}_{\alpha}(\vec{r}_{ij}){u}_{i,\beta} \delta(\vec{r}-\vec{r}_i) \;.
    \label{eq:kappa_m}
\end{gather}
\label{eq:dec_Im}
\end{subequations}

\begin{figure}
    \centering
    \includegraphics[]{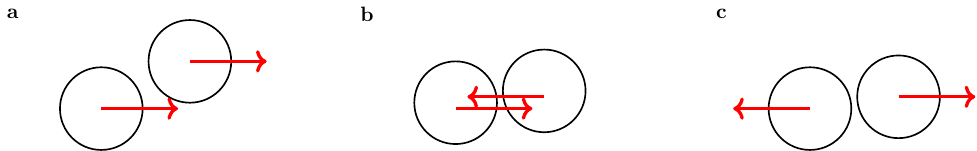}
    \caption{Depending on their relative orientations, pairs of particles provide varying contributions to the tensor ${\kappa}^{(m)}_{\alpha\beta}$ defined in Eq.~\eqref{eq:kappa_m}. ({\bf a}) Particles with parallel orientation provide a small contribution to ${\kappa}^{(m)}_{\alpha\beta}$. ({\bf b-c}) Particles with anti-parallel orientations provide larger contributions to $\hat{\kappa}^{(m)}_{\alpha\beta}$. However, we expect such contribution to be larger when two particles move towards one another ({\bf b}) compared to when they move away from each other ({\bf c}) as they tend to remain in contact for a longer time in the former situation.}
    \label{fig:sketch_coll}
\end{figure}

Then, we note that $\hat{\mathbf{\rm \kappa}}^{(m)}$ receives contributions of small magnitudes from pairs of particle with the same orientations, Fig.~\ref{fig:sketch_coll}a, and contributions of large magnitude from pairs of  particles with opposite orientations, see Fig.~\ref{fig:sketch_coll}b-c. Among pairs with opposite orientations, we expect a larger contribution when two particles are facing one another, as in Fig.~\ref{fig:sketch_coll}b, since repulsive force maintains them in contact for a longer time. These observations suggest that ${\rm \kappa}^{(m)}_{\alpha\beta}(\br) = \avg{\hat{\rm \kappa}^{(m)}_{\alpha\beta}(\br)}$ is related to the repulsion-induced reduction of the effective speed, i.e. ${\rm \kappa}^{(m)}_{\alpha\beta} \sim v^*[\rho]-v(\tilde\rho)$. We thus extend the ansatz of~\cite{SM-Omar_2023} to our density-dependent self-propulsion speed and assume:
\begin{equation}\label{eq-app:k-ansatz}
    \kappa^{(m)}_{\alpha\beta}(\br)= [v^*(\rho)-v(\tilde\rho)] \langle u_{\alpha} u_{\beta} \rangle = [v^*(\rho)-v(\tilde\rho)] (\rho \delta_{\alpha\beta}/2 + Q_{\alpha\beta}) \;,
\end{equation}
where $v^*(\rho)$ is the reduced local effective speed. As discussed in Sec.~\ref{sec:local}, this effective reduction is due to short-ranged collisions so that we write $v^*(\rho)=v(\tilde\rho)U(\rho)$. 
We then perform a mean-field approximation in Eq.~\eqref{eq:current_balance} and use the ansatz of Eq.~\eqref{eq-app:k-ansatz} to obtain
\begin{equation}
    0=\db \sigma^{\rm IK}_{\alpha\beta} - \frac{1}{D_r} v(\tilde\rho) \db \left[  v^*(\rho) \left(\frac{1}{2}\rho\delta_{\alpha \beta}+ {Q}_{\alpha \beta}\right) + \partial_\gamma\Sigma^{(m)}_{\alpha\beta\gamma} \right] \;.
    \label{smeq:inter_step}
\end{equation}
To close Eq.~\eqref{smeq:inter_step}, we need to compute the dynamics of the average nematic tensor $Q_{\alpha \beta}$ and evaluate it at steady state. To do so, we first multiply Eq.~\eqref{eq:fluct_dynamics} by $\left(u_{\alpha} u_{\beta} - \delta_{\alpha\beta}/2\right)$, integrate over the orientations $\theta$, and average over noise. All in all, this yields the following expression at steady state:
\begin{equation}
\label{eq-app:dynamics-Q}
    4D_r Q_{\alpha\beta} = - \partial_\gamma \left[ \avg{ \hat v \left( \hat B_{\alpha\beta\gamma} + \frac{1}{4}\eta_{\alpha\beta\gamma\nu} \hat m_\nu\right)} + \avg{\hat {\rm I}^{(Q)}_{\alpha\beta\gamma}}  - \frac{1}{2}\left( \langle \hat v \hat m_\gamma \rangle  + \partial_\nu \sigma^{\rm IK}_{\gamma\nu}\right)\delta_{\alpha\beta} \right]\;,
\end{equation}
where $\hat B_{\alpha \beta \gamma}$ is a third-order, symmetric traceless tensor representing the third-order moment of $\hat \psi$; $\eta_{\alpha \beta \gamma \nu} $ is defined as  $\eta_{\alpha \beta \gamma \nu} =\left[ \delta_{\alpha \beta} \delta_{\gamma \nu} + \delta_{\alpha \gamma} \delta_{\beta \nu} + \delta_{\alpha \nu} \delta_{\gamma \beta}\right] $, and 
\begin{equation}
   {\rm I}^{(Q)}_{\alpha\beta\gamma}= \avg{\hat {\rm I}^{(Q)}_{\alpha\beta\gamma}} = \avg{\sum_{ij}F_\gamma (\br_{ij})\left[ u_{i,\alpha}u_{i,\beta} -\delta_{\alpha \beta}/2 \right]\delta(\br-\br_i)} \;.
\end{equation}
To determine a closed expression for $Q_{\alpha\beta}$ we proceed as follows. First, in analogy with Eq.~\eqref{eq:dec_Im} for ${\rm I}^{(m)}_{\alpha\beta}$, we decompose ${\rm I}^{(Q)}_{\alpha\beta\gamma}$ as:
\begin{equation}
    {\rm I}^{(Q)}_{\alpha\beta\gamma}= \avg{\sum_{ij}F_\gamma (\br_{ij})(u_{i,\alpha}u_{i,\beta} - u_{j,\alpha}u_{j,\beta})\delta(\br-\br_i)} + \partial_\nu \Sigma^Q_{\alpha\beta\gamma\nu}\;,
\end{equation}
where $\partial_\nu \Sigma^Q_{\alpha\beta\gamma\nu}$ contributes to higher order in gradients and can thus be neglected. We then 
use the same ansatz as for ${\rm I}^{(m)}_{\alpha \beta}$ and rewrite ${\rm I}^{(Q)}_{\alpha\beta\gamma}$ as:
\begin{equation}
\label{eq:ansatz-Q}
    {\rm I}^{(Q)}_{\alpha \beta \gamma} = [v^*(\rho)-v(\tilde\rho)] ( B_{\alpha\beta\gamma} + \frac{1}{4}\eta_{\alpha\beta\gamma\nu} m_\nu) \;.
\end{equation}
We then inject Eq.~\eqref{eq:ansatz-Q} into Eq.~\eqref{eq-app:dynamics-Q}, and discard all contributions from the harmonic tensor $\hat B_{\alpha\beta\gamma}$, which are of higher order in the gradient expansion. Finally, in a flux-free steady state, $\langle \hat v m_\alpha\rangle + \partial_\beta \sigma^{\rm IK}_{\alpha\beta} = 0$, so that the last two terms in Eq.~\eqref{eq-app:dynamics-Q} also cancel out. All in all, Eq.~\eqref{eq-app:dynamics-Q} becomes:
\begin{align}
    Q_{\alpha\beta} &=  -\frac{1}{16D_r} \eta_{\alpha\beta\gamma\nu}\partial_\gamma \left[  \avg{\hat v m_\nu } - \avg{\hat v m_\nu} + \avg{\frac{v^*(\rho)}{\hat{v}} \hat{v} m_\nu}\right] + \cO(\nabla^4) \\[0.2cm]
    &= -\frac{1}{16D_r} \eta_{\alpha\beta\gamma\nu}\partial_\gamma \left[ \frac{v^*(\rho)}{v(\tilde\rho)}\partial_\lambda\sigma^{\rm IK}_{\nu\lambda} \right] + \cO(\nabla^4) \;,
\label{eq-app:Q-steady-state}
\end{align}
where in the last passage we used the mean-field approximation and the no-flux condition $\langle \hat v m_\nu\rangle = - \partial_\lambda \sigma^{\rm IK}_{\nu\lambda}$. We can now inject Eq.~\eqref{eq-app:Q-steady-state} into Eq.~\eqref{eq:current_balance} to obtain:
\begin{equation}
    0 = \da p^{\rm IK} + \frac{v(\tilde\rho)}{2D_r}\da[\rho v^*(\rho)] + \frac{v(\tilde\rho)}{16D_r^2}\partial_\beta\left[ v^*(\rho)\partial_\gamma\left( \frac{v^*(\rho)}{v(\tilde\rho)}\eta_{\alpha\beta\gamma\nu}\partial_\lambda \sigma^{\rm IK}_{\nu\lambda} \right) + 16D_r\partial_\gamma\Sigma^{(m)}_{\alpha\beta\gamma} \right] + \cO(\nabla^5)\;.\label{eq:closed_balance}
\end{equation}
To obtain the final expression to third order in the gradient expansion, we finally expand the non-local QS-contribution $v(\tilde\rho) = v(\rho) + \ell^2 v'(\rho) \nabla^2 \rho + \cO(\nabla^4)$, where $\ell^2 \equiv \frac{1}{2d} \int \rmd \br K(\br) \br^2$ and $K$ is the QS-interaction kernel. By retaining only the leading-order contributions, this finally yields:
\begin{equation}
    0 = \da p^{\rm IK} + \frac{v(\rho)+ \ell^2 v'(\rho) \nabla^2 \rho}{2D_r}\da[\rho v^*(\rho)] + \frac{v(\rho)}{16D_r^2}\partial_\beta\left[ v^*(\rho)\partial_\gamma\left( \frac{v^*(\rho)}{v(\rho)}\eta_{\alpha\beta\gamma\nu}\partial_\lambda \sigma^{\rm IK}_{\nu\lambda} \right) + 16D_r\partial_\gamma\Sigma^{(m)}_{\alpha\beta\gamma} \right]\;.\label{eq:closed_balance-final}
\end{equation}

Then, we consider a system that is phase separated along the $x$-axis and translationally invariant along $y$. To get a close equation we then follow~\cite{SM-Omar_2023} and discard the contributions from $\partial_\gamma \Sigma_{\alpha\beta\gamma}$. At this stage, we could proceed numerically to evaluate the binodals~\cite{SM-Solon_2018_Rmap} but we choose to make one more approximation and set $\ell^2 = 0$ to facilitate the determination of the binodals. This amounts to treating the QS interactions locally while retaining the non-local contributions from pairwise forces. Finally, we use our ansatz $v^*(\rho) = v(\rho) U(\rho)$ to rewrite Eq.~\eqref{eq:closed_balance} as:
\begin{equation}
0 = \dx g(x,[\rho]) \qquad \text{where} \qquad g(x,[\rho]) = \int^{\rho} \rmd s \frac{1}{v(s)} \frac{d p^{\rm IK}(s)}{ds}+ \frac{\rho v(\rho) U(\rho)}{2D_r} +\frac{3 v(\rho)}{16 D_r^2}U(\rho)\dx\left[U(\rho)\dx p^{\rm IK} \right]
    \label{eq-app:prediction-gradient}
\end{equation}
plays the role of an effective chemical potential for the non-local theory. We then notice that $g$ can be conveniently rewritten as:
\begin{equation}
    \label{eq:synthetic}
    g =  g_0(\rho) + \lambda(\rho) (\dx\rho)^2 - \kappa(\rho) \dx^2\rho \;,
\end{equation}
where we defined:
\begin{align}
    g_0(\rho) = \int^{\rho} ds \frac{p^{\rm IK}(s)'}{v(s)} + \frac{\rho v(\rho) U(\rho)}{2D_r} \;, \qquad
    \kappa(\rho) = \frac{3}{16D_r^2}v(\rho)^2\frac{dp^{\rm IK}}{d\rho}\;, \qquad
    \lambda(\rho) &= -\frac{3}{16D_r^2}v(\rho)\frac{d}{d\rho}\left(v(\rho)\frac{dp^{\rm IK}}{d\rho}\right)\;.
    \label{eq:coeffs}
\end{align}

As a consistency check, we note that the local theory, Eq.~(3) of the main text, is recovered upon discarding all terms of $\cO(\nabla^3)$. Also, the theory for phase coexistence of Ref.~\cite{SM-Omar_2023} is retrieved upon setting $v(\rho) = v_0$.

\subsection{Predictions for binodals from the gradient theory}
\label{sec:Rmapping}
In this section, we use our gradient theory, Eq.~\eqref{eq:synthetic}, to provide refined  predictions for the binodals. We remark that the method described below is also employed to predict the phase-separation binodals at large $\rf$ for the orange region of Fig.~2 of the main text, where quorum sensing does not play any role and $v(\rho) \simeq v_0$.

As the non-equilibrium chemical potential $g$ cannot be generically derived from a free energy, we follow recent methods~\cite{SM-Solon_2018_Rmap, SM-Omar_2023,SM-duan2025phase} that have been proposed to predict  phase separation via a suitable change of variables. The idea is to introduce a bijective mapping $\rho \to R$ by solving $\kappa R'' =
-(2\lambda+ \kappa')R'$~\cite{SM-Solon_2018_Rmap}, which leads to:
\begin{equation}
  R(\rho) = \int^\rho \rmd s \frac{1}{v(s)} \frac{\rmd p^{\rm IK}(s)}{\rmd s}\;.
  \label{eq:Rmapping}
\end{equation}
Such a simple expression is due to the choice of setting $\ell^2=0$. Otherwise, $R$ has to be determined numerically. This choice of $R$ then allows us to define  a generalized free energy $\cL[R(\br)]$ such that: $g(\br) =
\delta \cL/\delta R(\br)$. 

Knowing $R(\rho)$, we can express the local part of the chemical potential, $g_0(\rho)$, as a
function of the pseudo-density $R$. Integration of $g_0$ over $R$
then provides us with the local part of the pseudo-free energy $\cL[R(\br)]$, which
we denote by $\Psi(R)$. The phase-coexistence conditions are then obtained from a common-tangent construction on $\Psi(R)$, which yields the liquid and gas binodals in
the $R$-space, denoted by
$R_{\ell}, R_g$, through:
\begin{equation}
  g_0 (R_\ell) = g_0(R_g) \equiv \mathfrak{g}_0 \;, \qquad \Psi(R_\ell) - R_\ell \mathfrak{g}_0  = \Psi(R_g) - R_g \mathfrak{g}_0  \;.
  \label{eq:Rmapping-coex}
\end{equation}
By inverting the mapping $R_g, R_\ell \to \rho_g, \rho_\ell$, we
finally obtain the densities of the two coexisting phases.

\begin{figure}
    \centering\includegraphics[width=0.8\textwidth]{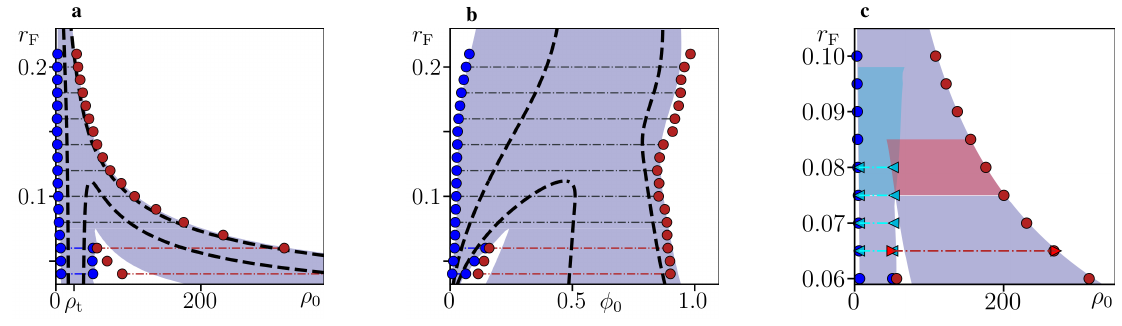}
    \caption{{Phase diagram predicted from the higher-order gradient theory}. \textbf{(a-b)} Phase diagram in the  $(\rho_0,\rf)$-plane {(a)} and in the $(\phi_0,\rf)$-plane {(b)} when $v(\rho\to\infty)=v_{\rm min}> 0$. Same simulation data as in Fig.~(3).
    Points connected by dash-dotted lines correspond to coexisting binodals obtained from simulations of Eq.~\eqref{eq-app:dynamics}. For $\rf \leq 0.06$, QS and PF binodals are depicted as blue and red circles, respectively. Dashed black lines indicate the spinodal lines, while the shaded region corresponds to the analytical prediction of the binodal region derived using our gradient theory of Eq.~\eqref{eq-app:prediction-gradient}. As expected, the gradient theory improves the predictions obtained via the local theory, which is plotted in Fig.~3.
    \textbf{(c)}  Metastable coexistence in the $(\rho_0,\rf)$-plane. Same data as in Fig.~4. Discs represent stable binodals; left-pointing cyan triangles stand for metastable binodals corresponding to QS-MIPS;  right-pointing red triangles stand for metastable binodals corresponding to PF-MIPS. Tie lines connect these coexisting metastable phases. Numerical results are compared to the analytical predictions for the binodals (shaded regions), obtained from Eq.~\ref{eq:Rmapping-coex}: stable regions are in blue, metastable QS-MIPS regions in cyan, and metastable PF-MIPS regions in red.
    }
    \label{fig:gradient_firstfig}
\end{figure}

In practice, we use the expressions of $U(\rho), p^{\rm IK}(\rho)$
derived in Section~\ref{sec:phase_coex} to compute $g, R$ and
the pseudo free-energy $\Psi$. We report our predictions as
shaded blue regions on the phase diagrams of
Fig.~\ref{fig:gradient_firstfig}, where they are compared to the data
points obtained from microscopic simulations.  As expected, the gradient theory brings a significant improvement in the accuracy of our predictions when compared to the local theory.

\if{
\subsection{Across the local approximations: A local theory is not preserved under a generic local transformation / change of variable}
(To me, this is a technical section that is not essential to understanding the paper but still is important for researchers in the field/wanting to replicate our theory. You all will know how to approach this better, so I'm just gonna try writing a first draft.)

From Eq.~\eqref{eq:1st_local_theory}, we might want a local theory $\partial_t \rho =\partial_\alpha [M(\rho)\partial_\alpha g(\rho)]$ with the factor $M(\rho)$ different from $v(\rho)$ in Eq.~\eqref{eq:1st_local_binodal}. The reasons could be to have an effective mobility satisfying the fluctuation-dissipation relation with the noise term that we averaged away, or to have a non-zero effective mobility in the case where the quorum-sensing speed goes to zero.

To change $M(\rho)$, the effective chemical potential would also change. Let the new chemical potential be $h(g(\rho))$. We can write, in one dimension,
\begin{equation*}
    M(\rho)\partial_x g(\rho) = \frac{M(\rho)}{h'(g(\rho))} \partial_x h(g(\rho))
\end{equation*}
Since this is a simple chain rule, the two common tangent constructions arising from these two theories should be equivalent. However, we find that this is not the case: a local theory transformed under $h$ is not preserved and gives a different common tangent construction. Below is an example illustrating that for a simple $h(g(\rho))=g(\rho)^2$.

(Insert figure with details of the math in the caption. The goal is to construct an example that is super easy to follow in one's head, so that it is very clear that this is real.)

This implies that the local theories are ill-defined and phase-separated profiles are necessarily stabilized by gradient terms.

Indeed, without a gradient term, the density profile of the phase-separated system is a step function, and the spatial derivative is a delta function centered at the interface, which breaks the derivation of the common tangent construction. (maybe go slower here)

Since the local theory is ill-defined, let's add the gradient terms back into the local approximation: $J_\alpha=M(\rho)\partial_\alpha [g(\rho) + \delta g[\rho]]$. Proceed as before:
\begin{align*}
    M(\rho)\partial_x g(\rho) &= \frac{M(\rho)}{h'(g(\rho)+\delta g[\rho])} \partial_x h(g(\rho)+\delta g[\rho]) \\
    &= \frac{M(\rho)}{h'(g(\rho)+\delta g[\rho])} \partial_x \left\{h(g(\rho))+h'(g(\rho))\delta g[\rho] \right\}
\end{align*}
The new chemical potential has the gradient term $h'(g(\rho))\delta g[\rho]$. If the original $\delta g[\rho]$ is equilibrium, this no longer will be, due to the term $h'(g)$. Therefore, we cannot use an equilibrium common tangent construction on the new theory. This explains why a local theory is not preserved under a generic local transformation. 

}\fi

\section{Absorbing phase transition: test of the kinetic argument and binodal predictions}
\label{sec:absorbing}
We first show the validity of the kinetic argument discussed in the main text for the absorbing phase transition. We then discuss a refined kinetic argument to account for the small deviations to $\rho_{\rm d}=2 \rho_{\rm t}$ observed in Figure~2. Finally, we discuss the derivation of the binodals reported in Fig.~2 for large $r_{\rm F}$.

\subsection{Test of the kinetic argument}
\label{sec:test}
\begin{figure}[htp!]
    \includegraphics[width=0.8\textwidth]{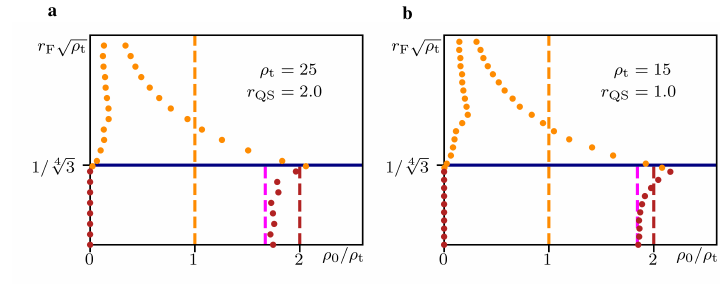}
    \caption{
    ({\bf a-b}) Phase diagram of the absorbing phase transition for different parameters. Red disks represent coexisting densities in the absorbing state, while orange disks represent coexisting densities in the liquid-gas region. The magenta dashed lines correspond to our refined prediction of the droplet density $\rho_{\rm d}$ at $r_{
    \rm F} \simeq 0$ given in Eq.~\eqref{eq:refined prediction}. (a) $\rho_{\rm t}=25, r_{\rm QS}=2, L_x=L_y=20$. The initial density is $\rho_0=16.4$ up to $\rf = 0.24$, where $\rho_0=16.4$ corresponds to a packing fraction $\phi_0\approx 0.65$. (Note that $\rf = 0.24$ is above the transition at $\bar{r}_{\rm F}=0.15$.) For larger $\rf$, $\rho_0$ is set such that $\phi_0\approx 0.65$, which ensures that the initial condition is between the MIPS binodals. (b) $\rho_{\rm t}=15, r_{\rm QS}=1, L_x=L_y=40$. The initial density is $\rho_0=4.7$ for $\rf < 0.44$. For $\rf \geq 0.44$, $\rho_0$ is chosen such that $\phi_0\approx 0.65$, for the same reason as in (a).  Other parameters correspond to the ones of Fig.~2 given in End Matter.
    }
    \label{fig:kinetic_arg}
\end{figure}

Let us show that the kinetic argument presented in the main text is robust to the choice of simulation parameters.
We consider systems where $v(\rho)$ vanishes beyond a threshold density $\rho_{\rm t}$. We thus expect the emergence of an absorbing dense phase where all particles arrest their motion when the pariwise-force interaction range $\rf$ is sufficiently small. Therefore, it exists a threshold radius $\bar r_{\rm F}$ below which the system condenses into arrested droplets. 
As detailed in the main text using a kinetic argument, we predict $\bar r_{\rm F}=(\sqrt{3}\rho_{\rm t})^{-1/2}$ below which we the system forms droplets of density $\rho_d \approx 2\rho_{\rm t}$. we remark that the predictions for $\rho_\rmd$ and $\bar r_{\rm F}$ do not depend on $\rq$.
To validate the kinetic argument we measure how the transition point $\bar{r}_{\rm F}$ and the arrested density $\rho_d$ change as we vary $\rho_{\rm t}$ and $r_{\rm QS}$. As shown in Fig.~\ref{fig:kinetic_arg} panels a and b, our predictions hold for numerical simulations with parameters different from those used in the main text. 

\subsection{Corrections to the droplet density due to the coarsening dynamics and the repulsive pairwise forces}
\label{sec:refined}
Let us now consider in more details the kinetics of the droplet growth and how it impacts its final density $\rho_{\rm d}$. We initialize our simulations in a homogeneous state with density $\rho_0= N/L^2$, which is small enough that we get a single arrested droplet in a simulation time of $T=1000$. Below, we first discuss the impact of $\rho_0$ on the droplet density in the limit of small $\rf$. 
We then show how, upon increasing $\rf$, compactification effects lead to an increase of $\rho_{\rm d}$, up to $\rho_{\rm d} = 2 \rho_{\rm t}$ at the transition point where $r_{\rm F}=\bar{r}_{\rm F}$.

We consider a spherical arrested cluster of radius $r$, surrounded by a gas of average density $\rho_{\rm g}(r)$. 
We assume that the gas is homogeneous and denote by $\rho_{\rm d}(r')$ the droplet density in a spherical shell $r'\in (r-\rmd r, r)$. We assume that (i) $r$ is large compared to $r_{\rm QS}$, (ii) $\rho_{\rm d}(r)$ varies over length scales large compared to $r_{\rm QS}$, and (iii) the effects of the droplet curvature can be neglected.

\begin{figure}[htb!]
    \centering
    \includegraphics[width=1.0\textwidth]{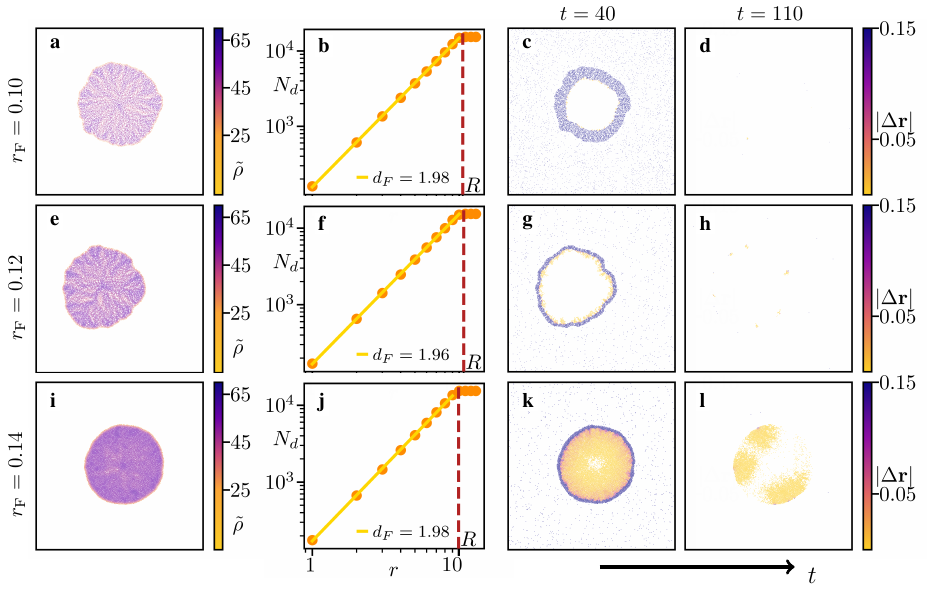}
    \caption{ Dynamics of droplet formation in the arrested phase, close to the transition point $\rf \lesssim \bar{r}_{\rm F}$. Each row corresponds to a simulation performed at different particle sizes $\rf$ at fixed quorum-sensing range $\rq=1$. \textbf{(a, e, i)}: Steady-state arrested droplet. Particles are colored based on the local density $\tilde\rho$. \textbf{(b, f, j)}: Number of particles $N_{\rm d}(r)$ inside a radius $r$ in log-log plot. The slopes of the lines of best fits are reported as the fractal dimension $d_F$. The vertical dashed lines are the maximum radii of the droplets. Distances are measured from the center of mass of the droplets. \textbf{(c, d, g, h, k, l)}: Dynamics of droplet formation, at early stage (time $t=40$, panels c,g,k) and late stage (time $t=110$, panels d,h,l), respectively. Particles are colored based on the local displacement $|\Delta \br|$ they undergo in the time window $[t-\Delta t, t)$, with $\Delta t = 10$; {particles are not plotted if they remained stationary, which we define as $|\Delta \br| < 10^{-5}$}. During droplet formation, incoming gas particles exert compressing forces on the droplet interface, resulting in its compactification. 
    Simulation parameters are the same as in Fig.~2.}
    \label{fig:snapshot_kinetic}
\end{figure}

Let us first discuss the case $r_{\rm F}\simeq 0$. For an incoming particle to be absorbed in the droplet, the total number of particles $N_{\rm local}$ in its surrounding must equal $\rho_{\rm t} \pi \rq^2$, so that its speed vanishes. The particle measures densities $\rho_{\rm d}(r)$  and $\rho_{g}(r)$ inside and outside the droplet, respectively, so that the arrest condition reads:
\begin{equation}
    \rho_{\rm t} = \frac{N_{\rm local}}{\pi \rq^2} = \frac{\rho_{\rm d}(r)+\rho_{\rm g}(r)}{2}  \;.
    \label{eq:arrest_Condition}
\end{equation}
To make progress, we note that the gas density $\rho_{\rm g}$ equals the number of particles that have not yet been absorbed by the droplet divided by the area occupied by the gas. 
Denoting $N_{\rm d}(r)$ the number of particles in the droplet, we have:
\begin{equation}
    \rho_{\rm g}(r) = \frac{N - N_{\rm d}(r)}{L^2 - \pi r^2} \;.
    \label{eq:rho_g_r}
\end{equation}
We note that, for all value of $r_{\rm F}$, Fig.~\ref{fig:snapshot_kinetic}(b,f,j) shows the droplet to have fractal dimension $d_F=2$ so that $N_{\rm d}(r) = 2 \pi \int_0^r \rho_{\rm d}(x) x \rmd x$. Together with Eq.~\eqref{eq:rho_g_r}, this allows eliminating $\rho_g$ to rewrite Eq.~\eqref{eq:arrest_Condition} as an ODE for $N_{\rm d}(r)$:
\if{
\begin{equation}
    \rho_{\rm d}(r) = 2 \rho_{\rm t} \left( \frac{\rq^2}{\rq^2 - \rf^2} \right) - \frac{N - 2 \pi \int_0^r \rho_{\rm d}(x) x \rmd x}{L^2 - \pi r^2} \;.
    \label{eq:refined_kinetic}
\end{equation}
}\fi

\begin{equation}
    \frac{N'_{\rm d}(r)}{2 \pi r} 
    =2 
    \rho_{\rm t}  - \frac{N - N_{\rm d}(r)}{L^2 - \pi r^2} \;.
\end{equation}
This equation can be readily solved to find
\begin{equation}
    N_{\rm d}(r) 
    = \frac{\pi r^2}{1-\pi r^2/L^2} \left[2  
    \rho_{\rm t} \left( 1 - \frac{\pi r^2}{2 L^2} \right) - \rho_0 \right] \qquad\text{and thus}\qquad \rho_{\rm d}(r) 
    =   \rho_{\rm t} +  \frac{ \rho_{\rm t} -\rho_0}{(1 - \pi r^2/L^2)^2} \;.
    \label{eq:Nd-rhod}
\end{equation}
Finally, the final size of the droplet $R$ corresponds to the whole system being absorbed, {\it i.e.}~$N_{\rm d}(R) = N$. We thus find:
\begin{equation}
    R^2 = 
    \frac{L^2}{\pi} \left( 1 - \sqrt{1 - \rho_0/{\rho}_{\rm t}}\right)\;,
    \label{eq:R2}
\end{equation}
which leads to $R \simeq 10$ for our choice of parameters, in agreement with our numerical observations, see Fig.~\ref{fig:snapshot_kinetic}(b,f,j). The \textit{average} density in the droplet then reads:
\begin{equation}
   \rho_{\rm d}  = 
    \frac{N}{\pi R^2} = \frac{\rho_0}{1 - \sqrt{1-\rho_0/{\rho}_{\rm t}}} \;. 
    \label{eq:droplet_density_final}
\end{equation}
For small $\rho_0$, this reduces to 
\begin{equation}\label{eq:refined prediction}
\rho_{\rm d}  = 2 \rho_{\rm t} - \rho_0/2\;, 
\end{equation}
which tends to $2\rho_{\rm t}$ as $\rho_0\to 0$.  

\begin{figure}[htp!]
    \includegraphics[]{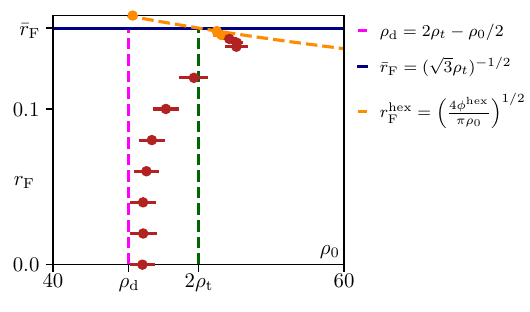}
    \caption{Droplet density in the arrested phase as a function of $\rf$ for the parameters of Fig.~2. Solid red disks correspond to arrested simulations, while orange disks correspond to liquid-gas coexistence without arrest. 
    The droplet density is estimated by binning the density field, building a density histogram, and fitting its peaks  with a Gaussian.
    Symbols and error bars correspond to the mean and standard deviation of the Gaussian fit, respectively. The orange dashed line corresponds to the value of $\rf$ such that the packing fraction equals $\phi^{\rm hex}$. At small $\rf$, Eq.~\eqref{eq:droplet_density_final} captures the droplet density. At higher $\rf$, the droplets undergo compactification due to incoming gas particles pushing on the particle at the interface. Finally, the droplet density decreases for $\rf \geq 0.14$, since compactification is limited by the maximal packing fraction $\phi_{\rm hex}$.
    }
    \label{fig:zoomed_droplet_density}
\end{figure}

Equation~\eqref{eq:droplet_density_final} agrees very well with our simulations at $r_{\rm F}=0$, as shown in Fig.~\ref{fig:zoomed_droplet_density}. As $\rf$ increases, we expect incoming particles to exert forces on the particles at the droplet interface, resulting in a droplet compactification and $\rho_{\rm d}>\rho_{\rm d}(r_{\rm F}=0)$. 
This is indeed what is observed in Fig.~\ref{fig:zoomed_droplet_density} as $r_{\rm F}$ increases away from $\rf=0$. 
However, the droplet density must remain below that of a hexagonal packing of discs of diameters $\rf$, which decreases with $\rf$ as $\rho^{\rm hex} = 4 \phi^{\rm hex} /(\pi \rf^2)$. 
As a result, the variations of $\rho_{\rm d}$ with $\rf$ eventually become non-monotonic as $\rf$ increases. The droplet density $\rho_{\rm d}$ thus decreases with $\rf$ close to the transition, as reported in Fig.~\ref{fig:zoomed_droplet_density}. The fact that the compactification of the droplet is due to the pressure exerted by incoming particles is visualized in Fig.~\ref{fig:snapshot_kinetic}(c,d,g,h,k,l), where we measure the particle net displacement $|\Delta \br|$ over a small time $\Delta t$. As $\rf$ increases, we observe a net particle displacement over a region of width much larger than $\rf$ near the droplet interface, leading to its compactification. 

\subsection{Binodal prediction for pairwise-forces MIPS}
\label{sec:MIPS}
To conclude this section, we explain how we predict the binodals in the region where the particle size $\rf$ is so large that QS interactions become negligible, i.e. the orange-shaded region in Fig.~2. In this region, the dense phase has a density $\rho_{\ell}$ such that $\rho_{\rm t}-\rho_\ell \gg \varphi$ in {$v_1(\rho)$} (see End Matter), so that all particles have approximately the same self-propulsion {$v_{\rm max}=v_0({\rm exp}(\lambda)-1)$}. We can thus neglect the dependence of $v$ on $\rho$ and follow the method of Sec.~\ref{sec:Rmapping} to compute the binodals via a common-tangent construction in the $R$-space. For $U(\phi)$ and $h(\phi)$ we use Eqs.~\eqref{eq:pressure-eq} with the same parameters as in Sec.~\ref{sec:phase_coex}. The lower boundary $\rf=\rf^{\rm PF}$ for this region is set when the predicted liquid density $\rho_\ell$ becomes equal to the QS threshold density $\rho_{\rm t}$.

\if{
This region is obtained by requiring that the dense-phase density $\rho_{\ell}\leq\rho_{\rm t}$, so that all particles have the same self-propulsion $v_{\rm max}$ independently of the local density. 
Phase separation is thus predicted following the methods of Ref.~\cite{SM-Omar_2023} which is equivalent to Eqs.~\eqref{eq:synthetic}---~\eqref{eq:coeffs} in the absence of QS, i.e. with $v(\rho)= v_{\rm max}$.
Binodals are constructed using the R-mapping and common-tangent construction presented in Sec.~\ref{sec:Rmapping}. 
As detailed in Sec.~\ref{sec:phase_coex}, we first need a closed expression for $U(\phi)$ and $h(\phi)$. To do so, we consider homogeneous systems of size $L_x=L_y=200$ for $\rf=1$, $v_0=v_2(\rho=0) = 5$, and $\ell_p^{\rm eff}=\frac{v}{\rf D_r}= 10$ as the initial density $\rho$ is varied. 
\qn{Make it more consistent with the EoS we determined above.}
We do a fit to the following functional form: 
\begin{equation}
    U(\phi) = (1-A\phi + B\phi^2) \frac{1-\tanh[C(\phi-D)]}{2}  \;, \qquad p^{\rm IK} (\rho,\rf) = \frac{v_{\rm max}r_{{\rm F,}0}}{\tilde v_0\rf} \left[ E\left(e^{F\phi}-1 \right) + G\left(e^{H\phi}-1 \right) \right]\;,
    \label{eq:pressure_pfaps}
\end{equation}
and we got $A = 1.23, B = 0.339, C=8.35, D=0.840, E=6.72\times 10^{-2}, F=5.51,G=5.92\times 10^{-9}, H=18.9$. These functions are then fed into the effective chemical potential $g$ and the $R$-mapping described by Eqs.~\eqref{eq:synthetic}---\eqref{eq:Rmapping}. We finally solve for the common tangent construction in the $R$ space, Eq.~\eqref{eq:Rmapping-coex}, and determine the resulting binodals.
}\fi

\if{
Not easy to get densification.\\

Scale of the effective potential is $V_0 \sim v_0 \alpha_1/\alpha_0$.\\
Effective temperature: $T \sim v_0^2/2\alpha_0$ so that the ratio between the two is $V_0/T \sim 2 \alpha_1/v_0$. Phase separation is obtained when this ratio is at $400$. \\

If we push too much $v_0$, the temperature raises and we lose phase separation.

If we increase $\alpha_1$ we go deeper in phase separation but we risk more negative tumbling rates. 

In Jeremy's paper they use $\alpha_0 = 100, \alpha_1 = 200, v_0 = 1$. Can we find a regime where $v_0$ is a bit larger, which would be convenient for PF phase separation? 

We'll see in the next episode.

However, with taxis on the self-propulsion via $v = v_0 - v_1 \mathbf{u} \cdot \nabla c$, the effective potential reads:
\begin{equation}
    U(x) = \frac{\gamma}{d} \left[ v_{1r} \frac{a_r}{\nu_r}G_r(x) - v_{1a} \frac{a_a}{\nu_a}G_a(x)\right]
\end{equation}
while the diffusivity reads, at zero temperature:
\begin{equation}
    D = \frac{v_0^2}{d \alpha_0} \;.
\end{equation}
As such, the ratio between energy and entropy in this latter case scales as:
\begin{equation}
    U/{S} \sim \frac{v_1}{D} \sim \frac{v_1}{v_0^2} \alpha_0 \sim \frac{v_1}{v_0} \frac{\alpha_0}{v_0}\;.
\end{equation}
We want this ratio to be $\sim 200$, $v_1/v_0 \sim 2$, $v_0/\alpha_0\sim 2$. This is pretty much impossible.

Mapping with previous problem:
\begin{equation}
    \frac{v_0 \alpha_1}{\alpha_0} \leftrightarrow v_1 \;.
\end{equation}
}\fi

\section{List of supplementary movies}\label{SMsec:movies}
In this section, we report all captions and parameters for the Supplementary Movies.

\begin{itemize}
    \item \textbf {SM Movie 1.} Absorbing phase transition for $\rf=0.14$. Same parameters as Fig. 1 panel a, see End Matter.

    \item \textbf {SM Movie 2.} Phase coexistence for $\rf=0.18$. Same parameters as Fig. 1 panel b, see End Matter.

    \item \textbf {SM Movie 3.} Phase coexistence for $\rf=0.04$. Same parameters as Fig. 1 panel c, see End Matter.

    \item \textbf {SM Movie 4.} Metastable coexistence that nucleates into a stable coexistence for $\rf=0.08$. Same parameters as Fig. 1 panel d and Fig. 4 panel e, see End Matter. We use the color code corresponding to the former.
\end{itemize}

\newpage

\end{document}